\documentclass[11pt,a4paper]{article}
\usepackage{graphicx}
\usepackage{amssymb}
\usepackage{amsmath}
\usepackage{amsfonts}
\usepackage{dsfont}
\usepackage{mathtools}
\usepackage{array}
\usepackage{subfig}
\usepackage{rotating}
\usepackage{bbold,amsfonts}
\usepackage{comment}
\usepackage[normalem]{ulem}
\usepackage{soul}

\usepackage{tikz}
\usetikzlibrary{decorations.markings}
\usetikzlibrary{arrows}

\usepackage[utf8]{inputenc}
\usepackage{bm}
\usepackage{xcolor}
\usepackage{float}
\usepackage{braket}
\usepackage{placeins}
\usepackage[height=8.8in,width=6.45in]{geometry}
\usepackage[font=small,labelfont=bf]{caption}
\usepackage[hidelinks]{hyperref}
\usepackage{booktabs,float,slashed}
\usepackage{standalone}
\usepackage{caption}
\usepackage{makecell}
\usepackage[maxbibnames=99,sorting=none,giveninits=true,backend=biber,style=numeric-comp,sortcites,doi=false,hyperref=true]{biblatex}
\renewbibmacro{in:}{}
\usepackage{hyperref}
\DeclareFieldFormat{doilink}{\iffieldundef{doi}{#1}{\href{https://doi.org/\thefield{doi}}{#1}}}
\DeclareFieldFormat[article,periodical]{volume}{\mkbibbold{#1}}
\DeclareFieldFormat[article,periodical]{journaltitle}{#1}
\DeclareFieldFormat[article,periodical]{pages}{#1}

\usepackage{xpatch}
\xpatchbibdriver{article}
  {\usebibmacro{journal+issuetitle}%
   \newunit
   \usebibmacro{byeditor+others}%
   \newunit
   \usebibmacro{note+pages}}
  {\printtext[doilink]{%
     \usebibmacro{journal+issuetitle}%
     \newunit
     \usebibmacro{byeditor+others}%
     \newunit
     \usebibmacro{note+pages}}}
  {}{}

\numberwithin{equation}{section}

\begin{document}

\begin{titlepage}
\vspace*{10mm}
\begin{center}
{\LARGE \bf 
Wilson loop correlators at strong coupling 

\vspace{0.2cm}

in $\mathcal{N}=2$ quiver gauge theories
}

\vspace*{15mm}

{\Large Alessandro Pini and Paolo Vallarino}

\vspace*{8mm}
	
 Universit\`a di Torino, Dipartimento di Fisica\\
 and I.N.F.N. - sezione di Torino
			
			\vskip 0.3cm
						
   Via P. Giuria 1, I-10125 Torino, Italy

\vskip 0.8cm
	{\small
		E-mail:
		\texttt{apini,vallarin@to.infn.it}
	}
\vspace*{0.8cm}
\end{center}

\begin{abstract}
We consider 4-dimensional $\mathcal{N} = 2$ superconformal quiver theories with $SU(N)^M$ gauge group and bi-fundamental matter and we evaluate correlation functions of $n$ coincident Wilson loops in the planar limit of the theory. Exploiting specific untwisted/twisted combinations of these operators and using supersymmetric localization, we are able to resum the whole perturbative expansion and find exact expressions for these correlators that are valid for all values of the 't Hooft coupling. Moreover,  we analytically derive the leading strong coupling behaviour of the correlators, showing that they obey a remarkable simple rule. Our analysis is complemented by numerical checks based on a Padé resummation of the perturbative series.
\end{abstract}
\end{titlepage}
\setcounter{tocdepth}{2}
\tableofcontents

\section{Introduction}
The study of the strong coupling regime in four dimensional gauge theories is a very challenging subject to tackle, yet, over the years, there have been considerable developments, especially in theories with a high amount of supersymmetry. Specifically many results have been achieved in the planar limit of the maximally supersymmetric theory in four dimensions, i.e. $\mathcal{N}=4$ Super Yang-Mills (SYM), by using supersymmetric localization, holography and integrability. However, more recently, mainly by exploiting supersymmetric localization techniques \cite{Pestun:2007rz} also the strong coupling regime of some $\mathcal{N}=2$ gauge theories in the large-$N$ limit has been investigated. In the context of  theories with eight supercharges many BPS observables have been analysed using localization techniques both at weak and strong coupling, such as correlation functions between chiral/anti-chiral scalar operators \cite{Gerchkovitz:2016gxx,Rodriguez-Gomez:2016ijh,Rodriguez-Gomez:2016cem,Fiol:2021icm,Bobev:2022grf,Beccaria:2021hvt,Billo:2022lrv,Billo:2022lrv,Billo:2022xas,Beccaria:2020hgy,Baggio:2016skg,Baggio:2014sna}, correlators between a chiral operator and a Wilson loop \cite{Sysoeva:2017fhr,Billo:2018oog,Galvagno:2021bbj,Preti:2022inu,Pini:2023svd}, the vacuum expectation value of BPS Wilson loops \cite{Billo:2019fbi,Beccaria:2021ksw,Beccaria:2021vuc} and the free energy \cite{Beccaria:2022kxy,Beccaria:2021ism,Fiol:2021jsc,Fiol:2020bhf}.

A particular class of $\mathcal{N}=2$ theories whose properties deserve to be analysed are the quiver gauge theories obtained by  a $\mathbb{Z}_M$ orbifold projection from $\mathcal{N} = 4$ SYM. These are superconformal quiver theories with $SU(N)^M$ gauge group and bi-fundamental matter hypermultiplets. Furthermore, they have a holographic dual realized by Type II B string theory on $AdS_5\times S^5/\mathbb{Z}_M$ space-time \cite{Kachru:1998ys,Gukov:1998kk,Lawrence:1998ja}, thus they represent one of the simplest models in which to investigate the holographic correspondence when supersymmetry is not maximal. In the framework of these quiver theories many exact results for different correlation functions in the large-$N$ limit have been found over the last few years \cite{Pini:2017ouj,Billo:2021rdb,Billo:2022gmq,Billo:2022fnb,Beccaria:2022ypy,Galvagno:2020cgq,Mitev:2014yba,Mitev:2015oty,Fiol:2020ojn}.

Moreover these quiver gauge theories are the starting point for the construction of other theories by means of orientifold projections (see e.g. \cite{Dey:2013fea}). For example from the $M=2$ quiver one obtains the so-called \textbf{E}-theory, which has  $SU(N)$ gauge group and matter in
the symmetric and anti-symmetric representations, that establishes another very fruitful playground where to find exact results and explore the strong coupling regime of gauge theories (see e.g. \cite{Beccaria:2020hgy,Beccaria:2021hvt,Pini:2023svd,Billo:2022xas}). 

Most of the strong coupling results obtained for BPS-observables in the quiver gauge theories have been carried out at the orbifold fixed point, the most symmetric configuration where all the Yang-Mills couplings associated to the nodes of the quiver are taken to be equal, i.e. $g_I \equiv g$. This way we introduce a unique 't Hooft coupling, namely $\lambda \equiv g^2N $.

Indeed, at the orbifold fixed point, a very useful tool that allows to get exact results for every value of the 't Hooft coupling in the planar limit of the theory becomes available. This is the $\textsf{X}$-matrix, a semi-infinite matrix whose elements are a convolution of Bessel functions of the first kind. As a matter of fact, as shown in \cite{Billo:2021rdb,Billo:2022gmq,Billo:2022lrv,Beccaria:2022ypy}, it is possible to express the partition function, and also 2- and 3- point functions of chiral scalar operators, in terms of this semi-infinite matrix. Furthermore, in recent papers \cite{Beccaria:2022ypy,Billo:2022lrv}  the features of the $\textsf{X}$-matrix have been employed to work out a systematic strong coupling expansion in inverse powers of the't Hooft coupling. 

In this article we focus on the study of  correlators among coincident Wilson loops in the fundamental representation in $\mathbb{Z}_M$ quiver gauge theories. Previous works on the same topic \cite{Rey:2010ry,Ouyang:2020hwd,Zarembo:2020tpf,Galvagno:2021bbj} considered the more general configuration where all the couplings are different. Although very interesting, this general set up makes more complicated the analysis of the theory at strong coupling. As a matter of fact the authors of \cite{Rey:2010ry,Zarembo:2020tpf,Ouyang:2020hwd} only focus on the v.e.v. of a single Wilson loop.
Instead, one of the main novelty of this work is that we only concentrate on the orbifold fixed point of the theory where, as mentioned above, the strong coupling regime of the theory can be successfully investigated using supersymmetric localization and the techniques of \cite{Beccaria:2022ypy,Billo:2022lrv}. Moreover, following the analysis carried out in \cite{Pini:2017ouj} for chiral/anti-chiral operators, it turns out to be convenient to introduce a new basis of Wilson loop operators, namely, generalizing \cite{Rey:2010ry,Galvagno:2021bbj}, we consider untwisted and twisted combinations of the Wilson loops associated to the different nodes of the quiver.

Specifically in this paper  we  derive an exact expression, valid for every value of the 't Hooft coupling, for the $n$-point Wilson loop correlator in the planar limit. Then,  generalizing to the case of $4d$ $\mathcal{N}=2$ $\mathbb{Z}_M$ quiver theory the techniques introduced in \cite{Belitsky:2020qir,Belitsky:2020qrm, Beccaria:2023kbl}, we find the leading order of its strong coupling expansion, which agrees with the corresponding result in $\mathcal{N}=4$ SYM up to a numerical proportionality factor.  To the best of our knowledge, this is the first example of an exact expression for a generic correlator among $n$ coincident Wilson loops in the planar limit of a four dimensional $\mathcal{N}=2$ superconformal gauge theory for which also  the strong coupling behaviour is analytically derived. 

This paper is organized as follows: in Section \ref{sec:TheTheory} we review the main features of the $4d$ $\mathcal{N}=2$ quiver gauge theory and we introduce the untwisted and twisted Wilson loop operators. In Section \ref{matrixmodel} we briefly recall the relevant aspects of supersymmetric localization and of the corresponding interacting matrix model, furthermore we review the notion of reducible correlator in the large-$N$ limit, that will be extensively used in the following sections. Then, as a warm-up example, in Section \ref{sec:2pt} we consider the first non-trivial correlator between twisted Wilson loops, i.e. the 2-point function. We formally derive an exact expression for this correlator in the planar limit and we determine the leading term of its strong coupling expansion. For the simplest $\mathbb{Z}_2$ quiver  we cross check our analytic prediction through a Padé resummation of the perturbative series. In Section \ref{sec: 3pt} we carry out a similar analysis for the 3-point function. Then in Section \ref{sec: npoint} we extend our results to the most generic reducible  $n$-point Wilson loop correlator, determining its large-$N$ expression and finding the leading term of its strong coupling expansion. Finally we wrap up in Section \ref{sec: conclusions} with some closing remarks. More technical details concerning the properties of the Bessel functions and the derivation of the strong coupling expansion are collected in three appendices.

\section{The \texorpdfstring{$4d$ $\mathcal{N}=2$}{} quiver gauge theory}
\label{sec:TheTheory}
In this section we introduce the theory and the observables that we are going to consider. Along the way we also fix the notation.

\begin{figure}
    \centering
    \includegraphics[scale=1.0]{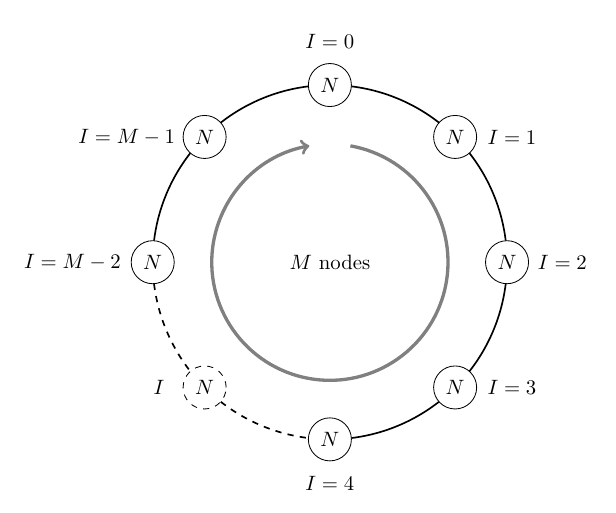}
    \caption{The circular quiver representing the $4d$ $\mathcal{N}=2$ SCFT with gauge group $SU(N) \times SU(N) \times\dots \times SU(N)$ and matter hypermultiplets in the bi-fundamental representation.}
    \label{fig:quiver}
\end{figure}

In this article we focus on the 4d $\mathcal{N}=2$ quiver gauge theory obtained by a $\mathbb{Z}_M$ orbifold of $\mathcal{N}=4$ SYM with gauge group $G=SU(NM)$. Its matter content is summarized by the quiver diagram reported in Figure \ref{fig:quiver},
where to each line between adjacent nodes is associated an $\mathcal{N}=2$ hypermultiplet transforming in the bi-fundamental representation $(\textbf{N},\overline{\textbf{N}})$ and, on the other hand, to each node of the quiver is associated an $\mathcal{N}=2$ vector multiplet transforming in the adjoint representation of the corresponding $SU(N)$ gauge group. In total there are $M$ nodes labeled by the index $I=0,\dots,M-1$.\footnote{All over the paper the index $I$ is taken modulo $M$, that is to say $I \sim I+M$.} This theory is conformal for arbitrary values of the coupling constants $g_I$, since there are $2N$ fundamental flavours at each $SU(N)$ node. Henceforth we only focus on the orbifold fixed point configuration, where all the couplings are taken to be equal, namely
\begin{align}
    g_I \equiv g \ \ \forall I \,.
\end{align}
Furthermore this theory has a non-trivial R-symmetry group given by $SU(2)\times U(1)_R$.

The gauge invariant operator that we are going to consider throughout this paper is the half-BPS Wilson loop along a circle $\mathcal{C}$ of radius $R$ in the fundamental representation of the $I$-th  node of the quiver \cite{Maldacena:1998im,Semenoff:2001xp}, namely
\begin{align}
\label{WilsonI} 
W^{(I)} \equiv \frac{1}{N}\,  \textrm{tr} \, \mathcal{P} \, \textrm{exp} \,  \Bigg\{ g_I \oint_{\mathcal{C}} \, d\tau \left[ i\,A^{I}_{\mu}\dot{x}^{\mu}(\tau) + \frac{R}{2}\left(\varphi^{I}(x)+\overline{\varphi}^{I}(x)\right)\right] \Bigg\} \,,
\end{align}
where $\mathcal{P}$ denotes the path-ordering, while $A_{\mu}^{I}$ and $\varphi^{I}$ the gauge field and the complex scalar of the $I$-th $\mathcal{N}=2$ vector multiplet. Finally, without any loss of generality, we place the circle $\mathcal{C}$ inside the $\mathbb{R}^2 \subset \mathbb{R}^4$ parameterized by $(x_1,x_2)$. Then the explicit expression of the function $x^{\mu}(\tau)$ reads
\begin{align}
\label{xmu}
x^{\mu}(\tau) = R\,(\cos(\tau), \sin(\tau), 0,0)\,  , \ \ \ \ \ 0 \leq \tau < 2\pi \, .
\end{align}
In this work we only consider circular Wilson loops with unitary radius, i.e. we set $R=1$.

Starting from \eqref{WilsonI} we define
\begin{subequations}
\begin{align}
& W_0 \equiv \frac{1}{\sqrt{M}}\left(W^{(0)}+W^{(1)}+\, \cdots \, + W^{(M-1)}\right)\,, \label{Wuntwisted} \\
& W_{\alpha} \equiv \frac{1}{\sqrt{M}}\sum_{I=0}^{M-1}\rho^{I\alpha}\,W^{(I)}\,, \label{Wtwisted}
\end{align}
\label{Wall}
\end{subequations}
where $\alpha=1,\dots, M-1$ and $\rho$ denotes the $M$-th root square of the identity, namely
\begin{align}
\label{rho}
\rho \equiv \text{e}^{2\pi i/M}\, .
\end{align}
The operator \eqref{Wuntwisted} is called untwisted Wilson loop, while the operators  \eqref{Wtwisted} are called twisted Wilson loops. Of course the two definitions \eqref{Wall} can be joined writing
\begin{align}
\label{Wfull}
W_{\alpha} \equiv \frac{1}{\sqrt{M}}\sum_{I=0}^{M-1}\rho^{I\alpha}\,W^{(I)}\, \ \, ,
\end{align}
where $\alpha=0,1,\dots,M-1$. 
\footnote{We recall that the notion of twisted/untwisted Wilson loop is not new. For instance the v.e.v. of a single Wilson loop at the orbifold fixed point was considered also in \cite{Galvagno:2021bbj,Rey:2010ry}, where it was shown that
\begin{align}
    \langle W_0 \rangle \simeq\ \sqrt{M}\langle W \rangle_0, \ \ \ \langle W_{\alpha} \rangle = 0\, \ ,
\end{align}
with $\alpha=1,\cdots,M-1$ and where $\langle W\rangle_0$ denotes the v.e.v. of the Wilson loop defined in $\mathcal{N}=4$.
}
The linear combinations \eqref{Wfull} turn out be very useful for the study of correlators among multiple coincident Wilson loops at the orbifold fixed point, that is
\begin{align}
\label{Wcorrelator}
\langle W_{\alpha_1}\,W_{\alpha_2}\dots W_{\alpha_n} \rangle \,.
\end{align}
Our goal is to examine the large-$N$ limit of \eqref{Wcorrelator}. Here we just notice that in order to get a non-trivial result one has to require
\begin{align}
\label{ZMchargetozero}
\sum_{i=1}^{n} \alpha_i = 0 \ \text{mod}  \ M\,.
\end{align}
Therefore, without loss of generality, we can always assume that $\alpha_n = -\alpha_1 -\alpha_2  \dots- \alpha_{n-1}$.\footnote{The origin of the relation \eqref{ZMchargetozero} can be easily understood by observing that, under a cyclic permutation $\varphi^{I} \mapsto \varphi^{I+1}$ that generates $\mathbb{Z}_M$, the operators \eqref{Wfull} transform as
\begin{align}
    W_{\alpha} \mapsto \rho^{\alpha}\,W_{\alpha} \, \ .
\end{align}
A generic $n$-point correlator of the type \eqref{Wcorrelator} must have charge zero under $\mathbb{Z}_M$. Therefore, the total twist of its operators must be zero modulo $M$ as expressed by the condition \eqref{ZMchargetozero}.} \\

Before addressing the most general case, we observe that the correlator of $n$ coincident untwisted Wilson loops in the $\mathbb{Z}_M$ quiver at the orbifold fixed point is planar equivalent to its counterpart in $\mathcal{N}=4$ SYM theory, namely in the 't Hooft limit it holds\footnote{Henceforth with $\simeq$ we denote the leading term in the large-$N$ expansion.}
\begin{align}
\langle \underbrace{W_{0}\,W_{0}\, \, \cdots \,W_{0}}_{n} \rangle\simeq  \bigl(\sqrt{M}\,\langle W \rangle_0\bigr)^n = \frac{2^n\,M^{n/2}}{\lambda^{n/2}}\bigl(I_1(\sqrt{\lambda})\bigr)^n\,.
\label{AllW0}
\end{align}
This expression can be regarded as an immediate extension of the results of \cite{Galvagno:2021bbj,Rey:2010ry,Ouyang:2020hwd}.
Since these correlators are planar equivalent to $\mathcal{N}=4$, in this work we focus on considering the cases of \eqref{Wcorrelator} with all $\alpha_i\neq 0$ or mixed correlators with both twisted and untwisted Wilson loops. 

In the next section we explain how the computation of the large-$N$ limit of \eqref{Wcorrelator} can be efficiently performed by exploiting supersymmetric localization.

\section{The matrix model}
\label{matrixmodel}
Using supersymmetric localization we can recast the computation of the expectation value of an observable on $\mathbb{R}^4$ in the evaluation of a finite dimensional matrix integral on the sphere $\mathbb{S}^4$. In this section we review the main properties of the matrix model for the 4d $\mathcal{N}=2$ quiver gauge theory introduced in Section \ref{sec:TheTheory} and we explain how, exploiting it, we can obtain exact expressions for the correlators \eqref{Wcorrelator} valid in the large-$N$ limit and for any value of the 't Hooft coupling.

\subsection{Explicit realization of the matrix model}
Once the theory is placed on a 4-sphere $\mathbb{S}^4$ of unitary radius, its partition function $\mathcal{Z}$ localizes and it can be written as a multiple integral over a set of $M$ matrices $a_I$ taking values in the $\mathfrak{su}(N)_I$ Lie-algebra
\begin{align}
\label{Zpartition}
\mathcal{Z} = \int \left(\prod_{I=0}^{M-1} \, da_I \right) \, \text{e}^{-\text{tr}\,a_I^2}\, |\mathcal{Z}_{1-loop}\,\mathcal{Z}_{inst}|^2 \, ,
\end{align}
where $\mathcal{Z}_{1-loop}$ contains the contributions due to the 1-loop determinants of the
fluctuations around the localization locus, while $\mathcal{Z}_{inst}$ encodes non-perturbative instanton corrections. However, since in the large-$N$ limit instantons are exponentially suppressed, we can ignore this contribution and henceforth we set $\mathcal{Z}_{inst} =1$. Here we use the so called ``Full-Lie algebra approach" (introduced in \cite{Billo:2017glv}) and therefore the integrations in \eqref{Zpartition} are performed over all the elements of the matrices $a_I$. We expand these matrices over the basis of the $\mathfrak{su}(N)_I$ generators $\{T_b\}$ writing
\begin{align}
a_I = a_{I}^bT_b\, \ \ \ \ \text{with} \ \ \ \  \text{tr}\,T_bT_c = \frac{1}{2}\delta_{b,c}\,,  
\end{align}
where $b,c=1,\dots, N^2-1$. This way the integration measure in \eqref{Zpartition} can be written as
\begin{align}
da_I = \prod_{b=1}^{N^2-1}\frac{da^{b}_I}{\sqrt{2\pi}}\,,
\end{align}
where we choose the normalization factor such that Gaussian integration over each $a_I$ is equal to $1$. Finally we observe that the 1-loop contribution can be rewritten in terms of an interaction action as
\begin{align}
|\mathcal{Z}_{1-loop}|^2 = \text{e}^{-S_{\mathrm{int}}} \,,
\end{align}
where the explicit expression of $S_{\mathrm{int}}$ is given by \cite{Billo:2021rdb}
\begin{align}
\label{Sint}
S_{\mathrm{int}} = \sum_{I=0}^{M-1}\left[\,
\sum_{m=2}^{+\infty}\sum_{k=2}^{2m}(-1)^{m+k}\Big(\frac{\lambda}{8\pi^2N}\Big)^{m}\,\binom{2m}{k}\,
\frac{\zeta_{2m-1}}{2m}\,\big(\textrm{tr} \, a_I^{2m-k}-\textrm{tr}\,a_{I+1}^{2m-k}\big) \big(\textrm{tr}\, a_I^k - \textrm{tr} \, a_{I+1}^k\big) \, \right]\,,
\end{align}
with $\zeta_{2m-1}$ denoting the odd Riemann $\zeta$-values. Thus the vacuum expectation value of any observable $f(a)$ can be written as
\begin{align}
\langle f(a) \rangle = \frac{\langle f(a) \, \text{e}^{-S_{int}} \rangle_0}{\langle \text{e}^{-S_{int}} \rangle_0} \,,
\end{align}
where $\langle \cdot \rangle_0$ stands for the v.e.v. in the free matrix model. The expectation values in the Gaussian matrix model can be efficiently evaluated by exploiting the set of identities (see equation (2.43) of \cite{Beccaria:2020hgy}) satisfied by 
\begin{align}
\label{tracesvev}
t_{n_1,\dots\, , n_p} \equiv \langle \text{tr}\,a^{n_1}\,\cdots \, \text{tr}\,a^{n_p} \rangle_0 \,,
\end{align}
where $a$ denotes a generic matrix in the $\mathfrak{su}(N)$ Lie-algebra. 

Now we want to show how we can express correlators of untwisted and twisted Wilson loops in the matrix model representation. In order to do that, we introduce the operators
\begin{align}
\label{Aoperators}
A_{\alpha,k}=\frac{1}{\sqrt{M}}\sum_{I=0}^{M-1}\rho^{\alpha I}\,\textrm{tr}\,a_I^k
\end{align}
with the understanding that
\begin{align}
A_{\alpha,k}^\dagger=A_{M-\alpha,k}\,.
\end{align}
Now, exploiting the definition of these operators, we can introduce the matrix model representation of the $1/2$ BPS circular untwisted and twisted Wilson loop of unit radius in the fundamental representation, namely \cite{Pestun:2007rz}
\begin{align}
\label{defWL}
W_{\alpha} = \frac{1}{\sqrt{M}\,N}\sum_{I=0}^{M-1}\rho^{\alpha I}\textrm{tr}\,\textrm{exp}\biggl[\sqrt{\frac{\lambda}{2N}}a_I \biggr]=\frac{1}{N}\sum_{k=0}^{\infty}\frac{1}{k!}\left(\frac{\lambda}{2N}\right)^{\frac{k}{2}}A_{\alpha,k} \, .
\end{align}
It is worth mentioning that the matrix model representation of the Wilson loop in $\mathcal{N}=4$ SYM is
\begin{align}
\label{defWLN=4} 
W = \frac{1}{N}\sum_{k=0}^{\infty}\,\frac{1}{k!} \left(\frac{\lambda}{2N}\right)^{\frac{k}{2}}\, \textrm{tr}a^{k}\,.
\end{align}
Therefore, through relation \eqref{defWL}, we connect the computation of Wilson loop correlators to that of correlation functions involving $A_{\alpha,k}$'s operators. From the definition \eqref{Aoperators} in general we have in the free matrix model
\begin{align}
\label{generalcorrA}
\langle A_{\alpha_1,k_1}A_{\alpha_2,k_2}\dots A_{\alpha_n,k_n} \rangle_0=\frac{1}{M^{n/2}}\sum_{I_1=0}^{M-1}\sum_{I_2=0}^{M-1} \dots \sum_{I_n=0}^{M-1}\rho^{-\alpha_1 I_1}\rho^{-\alpha_2 I_2}\dots \rho^{-\alpha_n I_n} \langle \textrm{tr}a_{I_1}^{k_1}\,\textrm{tr}a_{I_2}^{k_2}\,\dots \, \textrm{tr}a_{I_n}^{k_n} \rangle_0\,,
\end{align}
where $\sum_{i=1}^{n} \alpha_i = 0 \ \text{mod} \ M$. Specifically, one case that will be relevant in the following sections is given by the subset of \eqref{generalcorrA} with $\alpha_i=\alpha$ $\forall\,i=1,\dots ,M$. We find
\begin{align}
\label{generalconncorr}
    \langle A_{\alpha,k_1}A_{\alpha,k_2}\, \dots \, A_{\alpha,k_M} \rangle_0 = \frac{1}{M^{(M-2)/2}}\,t^{c}_{k_1,k_2,\dots,k_M}\,,
\end{align}
where $t^{c}_{k_1,k_2,\dots,k_M}$ is the connected correlator \footnote{The generating functional of the connected correlators is the free energy $\mathcal{F}=-\textrm{log}\,\mathcal{Z}_{\mathcal{N}=4}$, where $\mathcal{Z}_{\mathcal{N}=4}$ is the $\mathcal{N}=4$ partition function deformed by adding the sources $\{h_{k_i,I}\}$ for all the trace operators and is defined as
\begin{align}
\mathcal{Z}(\{h_{k_i,I}\}) = \int \left(\prod_{I=0}^{M-1} \, da_I \right) \, \text{e}^{-\text{tr}\,a_I^2-\sum_{i=1}h_{k_i,I}\textrm{tr}a_I^{k_i}}\,.
\end{align}
From the free energy is then possible to obtain the connected $n$-point functions as 
\begin{align}
t^c_{k_1,k_2,\dots,k_n}=\frac{\partial}{\partial h_{k_1,I}}\frac{\partial}{\partial h_{k_2,I}}\dots \frac{\partial}{\partial h_{k_n,I}}\mathcal{F}(\{h_{k_i,I}\})\Big|_{\{h_{k_i,I}=0\}}\,.
\end{align}
} in the Gaussian matrix model, i.e. in $\mathcal{N}=4$ SYM. For instance, for connected 2- and 3- point correlation functions in the free matrix model we have
\begin{align}
& t_{k,\ell}^c=t_{k,\ell}-t_kt_{\ell} \,, \label{2ptconn}\\
& t_{k,\ell,p}^c=t_{k,\ell,p}-t_kt_{\ell,p}-t_{\ell}t_{k,p}-t_pt_{k,\ell}+2\,t_kt_{\ell}t_p \label{3ptconn}\,,
\end{align}
where we have used the definition \eqref{tracesvev}. 

However, more importantly, we are interested in the large-$N$ behaviour of Wilson loop correlators and, for this purpose, as seen in \cite{Pini:2023svd}, it is convenient to introduce the vevless basis of the $A_{\alpha,k}$'s operators, i.e.
\begin{align}
\label{Ahatoperators}
\hat{A}_{\alpha,k}\equiv A_{\alpha,k}-\langle A_{\alpha,k} \rangle \,,
\end{align}
where from definition \eqref{Aoperators} it follows that
\begin{align}
\label{Atwisted}
\hat{A}_{\alpha,k}=A_{\alpha,k}\ \ \ \ \text{for}\ \ \alpha\neq 0\,.
\end{align}
As it has been shown in \cite{Billo:2021rdb,Billo:2022fnb}, it holds that
\begin{subequations}
\begin{align}
\label{Wick2A}
& \langle \, \hat{A}_{\alpha,k}\,\hat{A}_{\beta,\ell}^{\dagger}\, \rangle_0  \, \propto \, N^{\frac{k+\ell}{2}}\delta_{\alpha,M-\beta} \ \ \text{with} \ \ k+\ell \ \  \text{even}\,, \\[0.5em]
& \langle\, \hat{A}_{\alpha,k}\,\hat{A}_{\beta,\ell}\,\hat{A}_{\gamma,q}^{\dagger} \,\rangle_0 \, \propto \, N^{\frac{k+\ell+q}{2}-1}\delta_{\alpha+\beta,M-\gamma}\, , \ \text{with} \ k+\ell+q \, \ \ \text{even}\,.
\label{Wick3A}
\end{align}
\end{subequations}
Then the leading order of the large-$N$ expansion of a generic higher point correlator can be factorized à la Wick among the product of 2-point functions (for correlators involving an even number of $\hat{A}_{\alpha,k}$'s)  or the product of just one 3-point function and 2-point functions (for correlators involving an odd number of $\hat{A}_{\alpha,k}$'s). For example for a 4-point correlator we find
\begin{align}
\label{4ptA}
& \langle \hat{A}_{\alpha_1,k_1}\,\hat{A}_{\alpha_2,k_2}\,\hat{A}_{\alpha_3,k_3}\,\hat{A}_{\alpha_4,k_4}^{\dagger}\, \rangle_0 \simeq \langle\, \hat{A}_{\alpha_1,k_1}\,\hat{A}_{\alpha_2,k_2}^{\dagger}\,\rangle_0\langle\, \hat{A}_{\alpha_3,k_3}\,\hat{A}_{\alpha_4,k_4}^{\dagger}\,\rangle_0  \nonumber \\[0.5em]
& + \langle\, \hat{A}_{\alpha_1,k_1}\,\hat{A}_{\alpha_3,k_3}^{\dagger}\,\rangle_0\langle\, \hat{A}_{\alpha_2,k_2}\,\hat{A}_{\alpha_4,k_4}^{\dagger}\,\rangle_0  + \langle\,\hat{A}_{\alpha_1,k_1}\,\hat{A}_{\alpha_4,k_4}^{\dagger}\,\rangle_0\langle\, \hat{A}_{\alpha_2,k_2}\,\hat{A}_{\alpha_3,k_3}^{\dagger}\,\rangle_0 \,,  
\end{align}
where $k_1+k_2+k_3+k_4$ is even. We notice that in general  the coefficient of the leading term of the large-$N$ expansion could vanish and then the correlator would scale as $N^{\frac{k_1+k_2+k_3+k_4}{2}-1}$. If this turns out to be the case, the corresponding correlator is called irreducible since it cannot be factorized into the product of 2-point correlators, otherwise is called reducible.  An  example of irreducible correlator is provided by the 4-point function $\langle A_{1,k_1}\,A_{1,k_2}\,A_{1,k_3}\,A_{1,k_4} \rangle_0$ of the $\mathbb{Z}_4$ quiver gauge theory. We would like to warn the reader that, only in some specific cases (as in the example above), it is possible to determine a priori whether a correlator is irreducible. In fact, we are not aware of a general algorithmic rule that allows one to determine whether a correlator is irreducible solely by inspecting the list $\{\alpha_1, \alpha_2, \dots, \alpha_n\}$ of its twisted label. This is for sure a very interesting open problem that, however, lies beyond the scope of this article.\footnote{Of course, the same observation applies to correlators involving $n$-point coincident Wilson loops $\langle W_{\alpha_1}\,W_{\alpha_2}\,\dots\, W_{\alpha_n} \rangle$. Even in this case, there is no general procedure to determine whether such correlators are irreducible.} \\

\subsection{The \texorpdfstring{$\mathcal{P}$}{}-basis and its properties}
Although correlation functions of Wilson loops can be naturally expressed using the $\hat{A}_{\alpha,k}$ basis, in the interacting theory it is useful to use a new basis, the so called $\mathcal{P}_{\alpha,k}$ basis, that enormously simplifies the computations. This basis has been introduced in \cite{Billo:2022fnb} and here we just review its main properties. In the large-$N$ limit the $A_{\alpha,n}$ basis and the $\mathcal{P}_{\alpha,n}$ basis are related as follows
\begin{align}
\label{fromAtoP}
\hat{A}_{\alpha,k} \simeq \left(\frac{N}{2}\right)^{\frac{k}{2}}\sum_{i=0}^{\lfloor\frac{k-1}{2}\rfloor}\sqrt{k-2i}\left(\begin{array}{c}
     k  \\
     i 
\end{array}\right)\mathcal{P}_{\alpha,k-2i} \,.
\end{align}
The interaction action in \eqref{Sint} has a remarkable simple form in the $\mathcal{P}$-basis
\begin{align}
\label{SintPbasis}
S_{\mathrm{int}} = -\frac{1}{2}\sum_{\alpha=0}^{M-1}s_{\alpha}\,\mathcal{P}_{\alpha}^{\dagger}\,\textsf{X}\,\mathcal{P}_{\alpha} \,,
\end{align}
where $\mathcal{P}_{\alpha}$ is an infinite vector with entries $\mathcal{P}_{\alpha,k}$,
\begin{align}
\label{salfa}
s_{\alpha} \equiv \sin\left(\frac{\pi\alpha}{M}\right)^2 \,,
\end{align}
and $\textsf{X}$ is a semi-infinite symmetric matrix, whose entries with opposite parity vanish
\begin{align}
\label{XEvenOdd}
\textsf{X}_{2k,2\ell+1} \equiv 0 \,,
\end{align}
while its non-trivial elements read
\begin{align}
\label{Xmatrix}
\textsf{X}_{k,\ell} \equiv -8\,(-1)^{\frac{k+\ell+2k\ell}{2}}\,\sqrt{k\,\ell}\,\int_0^{\infty} \frac{dt}{t} \frac{\text{e}^{t}}{(\text{e}^{t}-1)^2}\,J_k\left(\frac{t\sqrt{\lambda}}{2\pi}\right)J_{\ell}\left(\frac{t\sqrt{\lambda}}{2\pi}\right) \,,
\end{align}
where $k,\ell \geq 2$.

Using the above expression for the interacting action \eqref{SintPbasis} one can show that the interacting 2-point function reads \cite{Billo:2022fnb}
\begin{align}
\label{2pointP}
& \langle\, \mathcal{P}_{\alpha,k}\,\mathcal{P}_{\beta,\ell}^{\dagger}\, \rangle \simeq \delta_{\alpha,\beta}\,\textsf{D}_{k,\ell}^{(\alpha)} \,,
\end{align}
where
\begin{align}
\label{Dalfa}
\textsf{D}_{k,\ell}^{(\alpha)} = \delta_{k,\ell} + s_{\alpha}\,\textsf{X}_{k,\ell} + s_{\alpha}^2\,\textsf{X}^2_{k,\ell} + \dots = \left(\frac{1}{1-s_{\alpha}\,\textsf{X}}\right)_{k,\ell} \,.
\end{align}
We notice that for the untwisted sector ($\alpha=0$) $\textsf{D}_{k,\ell}^{(0)}$ reduces to $\delta_{k,\ell}$.

The interacting 3-point function reads 
\begin{align}
\label{3pointP}
& \langle\, \mathcal{P}_{\alpha,k}\,\mathcal{P}_{\beta,\ell}\,\mathcal{P}_{\gamma,q}^{\dagger}\, \rangle \simeq \delta_{\alpha+\beta,\gamma}\,\frac{\textsf{d}_{k}^{(\alpha)}\textsf{d}_{k}^{(\beta)}\textsf{d}_{k}^{(\gamma)}}{M\sqrt{N}} \, ,
\end{align}
where
\begin{align}
\label{dalfa}  
\textsf{d}^{(\alpha)}_{k} = \sum_{\ell=2}^{\infty}\textsf{D}_{k,\ell}^{(\alpha)}\,\sqrt{\ell} \, .
\end{align}
Higher point functions in the interacting theory can be still computed using Wick's theorem. This is due to the fact that, at the leading term of the large-$N$ expansion, the $\mathcal{P}_{\alpha,k}$ operators inherit the properties \eqref{Wick2A}-\eqref{Wick3A} valid for the $\hat{A}_{\alpha,k}$ in the free theory. This way a correlation function among an even number of $\mathcal{P}_{\alpha,k}$ can be decomposed into the sum over all the possible Wick's contractions using the propagator \eqref{2pointP}. For example it holds that   
\begin{align}
\label{Example4point}
& \langle \,\mathcal{P}_{\alpha_1,k_1}\,\mathcal{P}_{\alpha_2,k_2}\,\mathcal{P}_{\alpha_3,k_3}\,\mathcal{P}_{\alpha_4,k_4}^{\dagger}\, \rangle \simeq \langle\, \mathcal{P}_{\alpha_1,k_1}\,\mathcal{P}_{\alpha_2,k_2}^{\dagger}\,\rangle\langle\, \mathcal{P}_{\alpha_3,k_3}\,\mathcal{P}_{\alpha_4,k_4}^{\dagger}\,\rangle  \nonumber \\[0.5em]
& + \langle\, \mathcal{P}_{\alpha_1,k_1}\,\mathcal{P}_{\alpha_3,k_3}^{\dagger}\,\rangle\langle\, \mathcal{P}_{\alpha_2,k_2}\,\mathcal{P}_{\alpha_4,k_4}^{\dagger}\,\rangle  + \langle\,\mathcal{P}_{\alpha_1,k_1}\,\mathcal{P}_{\alpha_4,k_4}^{\dagger}\,\rangle\langle\, \mathcal{P}_{\alpha_2,k_2}\,\mathcal{P}_{\alpha_3,k_3}^{\dagger}\,\rangle \, \ . 
\end{align}
On the other hand a correlator among an odd number of $\mathcal{P}_{\alpha,k}$ can be decomposed into the sum over all the possible Wick's contractions using the 2-point functions \eqref{2pointP} and just one 3-point function \eqref{3pointP}. For example
\begin{align}
& \langle\, \mathcal{P}_{\alpha_1,k_1}\,\mathcal{P}_{\alpha_2,k_2}\,\mathcal{P}_{\alpha_3,k_3}\,\mathcal{P}_{\alpha_4,k_4}\,\mathcal{P}_{\alpha_5,k_5}^{\dagger} \rangle \, \simeq \, \langle \,\mathcal{P}_{\alpha_1,k_1}\,\mathcal{P}_{\alpha_2,k_2}\,\mathcal{P}_{\alpha_3,k_3}^{\dagger}\, \rangle\,\langle \,\mathcal{P}_{\alpha_4,k_4}\,\mathcal{P}_{\alpha_5,k_5}^{\dagger}\, \rangle \nonumber\\[0.5em] & + \langle \,\mathcal{P}_{\alpha_1,k_1}\,\mathcal{P}_{\alpha_2,k_2}\,\mathcal{P}_{\alpha_4,k_4}^{\dagger}\, \rangle\,\langle \,\mathcal{P}_{\alpha_3,k_3}\,\mathcal{P}_{\alpha_5,k_5}^{\dagger}\, \rangle +\langle \,\mathcal{P}_{\alpha_1,k_1}\,\mathcal{P}_{\alpha_2,k_2}\,\mathcal{P}_{\alpha_5,k_5}^{\dagger}\, \rangle\,\langle \,\mathcal{P}_{\alpha_3,k_3}\,\mathcal{P}_{\alpha_4,k_4}^{\dagger}\, \rangle  \nonumber\\[0.5em]
& + \langle \,\mathcal{P}_{\alpha_1,k_1}\,\mathcal{P}_{\alpha_3,k_3}\,\mathcal{P}_{\alpha_4,k_4}^{\dagger}\, \rangle\,\langle \,\mathcal{P}_{\alpha_2,k_2}\,\mathcal{P}_{\alpha_5,k_5}^{\dagger}\, \rangle + \langle \,\mathcal{P}_{\alpha_1,k_1}\,\mathcal{P}_{\alpha_4,k_4}\,\mathcal{P}_{\alpha_5,k_5}^{\dagger}\, \rangle\,\langle \,\mathcal{P}_{\alpha_2,k_2}\,\mathcal{P}_{\alpha_3,k_3}^{\dagger}\, \rangle  \nonumber \\[0.5em]
& + \langle \,\mathcal{P}_{\alpha_1,k_1}\,\mathcal{P}_{\alpha_3,k_3}\,\mathcal{P}_{\alpha_5,k_5}^{\dagger}\, \rangle\,\langle \,\mathcal{P}_{\alpha_2,k_2}\,\mathcal{P}_{\alpha_4,k_4}^{\dagger}\, \rangle + \langle \,\mathcal{P}_{\alpha_3,k_3}\,\mathcal{P}_{\alpha_4,k_4}\,\mathcal{P}_{\alpha_5,k_5}^{\dagger}\, \rangle\,\langle \,\mathcal{P}_{\alpha_1,k_1}\,\mathcal{P}_{\alpha_2,k_2}^{\dagger}\, \rangle \nonumber \\[0.5em]
& + \langle \,\mathcal{P}_{\alpha_2,k_2}\,\mathcal{P}_{\alpha_4,k_4}\,\mathcal{P}_{\alpha_5,k_5}^{\dagger}\, \rangle\,\langle \,\mathcal{P}_{\alpha_1,k_1}\,\mathcal{P}_{\alpha_3,k_3}^{\dagger}\, \rangle +  \langle \,\mathcal{P}_{\alpha_2,k_2}\,\mathcal{P}_{\alpha_3,k_3}\,\mathcal{P}_{\alpha_5,k_5}^{\dagger}\, \rangle\,\langle \,\mathcal{P}_{\alpha_1,k_1}\,\mathcal{P}_{\alpha_4,k_4}^{\dagger}\, \rangle \nonumber \\[0.5em]
& + \langle \,\mathcal{P}_{\alpha_2,k_2}\,\mathcal{P}_{\alpha_3,k_3}\,\mathcal{P}_{\alpha_4,k_4}^{\dagger}\, \rangle\,\langle \,\mathcal{P}_{\alpha_1,k_1}\,\mathcal{P}_{\alpha_5,k_5}^{\dagger}\, \rangle \, \ . 
\label{Example5point}
\end{align}
We stress that the two relations \eqref{Example4point}-\eqref{Example5point} are valid in the large-$N$ limit and are exact in $\lambda$.

\section{The 2-point function \texorpdfstring{$\langle 
\label{sec:2pt}
W_{\alpha}W^{\dagger}_{\alpha} \rangle$}{}}
\label{sec:WW}
Let us begin the computation of Wilson loop correlators in the $\mathbb{Z}_{M}$ quiver gauge theory in the 't Hooft limit, starting from the simplest case, namely the 2-point function. The case of a 2-point function with one untwisted and one twisted Wilson loops is vanishing due to the $\mathbb{Z}_{M}$ symmetry of the quiver; we focus on the  case of a twisted 2-point function ($\alpha \neq 0$). Using the expression \eqref{defWL} we get
\begin{align}
\label{eq:doubleWW}
\langle W_{\alpha}W_{\alpha}^{\dagger} \rangle  = \frac{1}{N^2}\sum_{n=0}^{\infty}\sum_{m=0}^{\infty}\frac{1}{n!\,m!}\left(\frac{\lambda}{2N}\right)^{\frac{n+m}{2}}\langle A_{\alpha,n}\,A_{\alpha,m}^{\dagger} \rangle \,.
\end{align}
Since we are considering $\alpha\neq 0$, we can replace the $A_{\alpha,n}$ operators with their corresponding $\hat{A}_{\alpha,n}$ ones and then we can exploit the relation \eqref{fromAtoP} and  \eqref{2pointP} to write the leading term of the large-$N$ expansion of the correlator as
\begin{align}
\langle \hat{A}_{\alpha,n}\, \hat{A}_{\alpha,m}^{\dagger} \rangle \simeq \left(\frac{N}{2}\right)^{\frac{n+m}{2}}\sum_{i=0}^{\lfloor \frac{n}{2} -1 \rfloor}\,\sum_{j=0}^{\lfloor \frac{m}{2} -1 \rfloor}\left(\begin{array}{c}
     n  \\
     i 
\end{array}\right)\left(\begin{array}{c}
     m  \\
     j 
\end{array}\right)\sqrt{n-2i}\sqrt{m-2j}\,\textsf{D}^{(\alpha)}_{n-2i,m-2j} \,.
\end{align}
At this stage we consider the following redefinition of the indices
\begin{subequations}
\begin{align}
\label{redef}
& n-2i\mapsto k \,, \\
& m-2j\mapsto \ell \,,
\end{align}
\end{subequations}
therefore in the planar limit the expression \eqref{eq:doubleWW} becomes
\begin{align}
\label{2ptquiver}
\langle W_{\alpha}W_{\alpha}^{\dagger} \rangle \simeq \frac{1}{N^2}\sum_{k=2}^{\infty}\sum_{\ell=2}^{\infty}I_{k}(\sqrt{\lambda})I_{\ell}(\sqrt{\lambda})\sqrt{k\,\ell}\,\textsf{D}^{(\alpha)}_{k,\ell}\,,   
\end{align}
where we have used the expansion of the modified Bessel functions of the first kind
\begin{align}
\label{besselI}
I_{p}(x)=\sum_{i=0}^{\infty} \frac{1}{i!(i+p)!} \biggl( \frac{x}{2} \biggr)^{2i+p} \,.
\end{align}
Equation \eqref{2ptquiver} is the expression for the 2-point correlator valid for each value of the 't Hooft coupling in the planar limit of the quiver $\mathbb{Z}_M$. We notice that \eqref{2ptquiver} is exponentially divergent for $\lambda >> 1$, for this reason we find convenient to consider its ratio  with the corresponding observable in $\mathcal{N}=4$ SYM. This latter is given by
\begin{align}
\label{eq:doubleWWN=4}
\langle W_{\alpha}W_{\alpha}^{\dagger} \rangle_0  = \frac{1}{N^2}\sum_{n=0}^{\infty}\sum_{m=0}^{\infty}\frac{1}{n!\,m!}\left(\frac{\lambda}{2N}\right)^{\frac{n+m}{2}}\langle \hat{A}_{\alpha,n}\,\hat{A}_{\alpha,m}^{\dagger} \rangle_0 = \frac{1}{N^2}\sum_{n=0}^{\infty}\sum_{m=0}^{\infty}\frac{1}{n!m!}\left(\frac{\lambda}{2N}\right)^{\frac{n+m}{2}}t^{c}_{n,m} \,,
\end{align}
where $t^c_{k,\ell}$ is defined in \eqref{2ptconn}. Using the expression for the $\mathcal{N}=4$ Wilson loop in \eqref{defWLN=4}, we can rewrite this result as
\begin{align}
\label{Wconnected}
  \langle W_{\alpha}W_{\alpha}^{\dagger} \rangle_0  =  \langle W\,W \rangle_{0}- \langle W \rangle_{0}^2 \equiv   W_{conn}^{(2)}(\lambda) \,,
\end{align}
where $W_{conn}^{(2)}(\lambda)$ is the connected Wilson loop 2-point function that was previously considered also in \cite{Okuyama:2018aij}.
In the large-$N$ expansion one obtains\footnote{See Appendix \ref{appendix:A} for the details concerning this computation.} 
\begin{align}
\label{eq:Wcon}
 W_{conn}^{(2)}(\lambda) \simeq \frac{1}{N^2}\sum_{k=2}^{\infty}\sum_{\ell=2}^{\infty}\,I_{k}(\sqrt{\lambda})\,I_{\ell}(\sqrt{\lambda})\, \sqrt{k\, \ell}\, \delta_{k,l} = \frac{1}{N^2}\sum_{k=2}^{\infty} \, I_{k}(\sqrt{\lambda})^2\, k = \frac{\sqrt{\lambda}}{2N^2}I_{1}(\sqrt{\lambda})I_{2}(\sqrt{\lambda})  \,.
\end{align}
Therefore, we finally consider the ratio 
\begin{align}
\label{eq:ratioWW}
\frac{\langle W_{\alpha}W_{\alpha}^{\dagger}\ \rangle}{ W_{conn}^{(2)}(\lambda)} \equiv 1 +\Delta w^{(\alpha)}(M,\lambda) \,.
\end{align}
We observe that when we turn off the interaction action $S_{\mathrm{int}}$ the ratio \eqref{eq:ratioWW} is just equal to $1$.  In the following, we aim to study the leading order of the large-$\lambda$ expansion of 
$\Delta w^{(\alpha)}(M,\lambda)$ which, in the planar limit and for any value of the 't Hooft coupling, reads \footnote{We would like to warn the reader that the function $\Delta w^{(\alpha)}(M,\lambda)$ depends on $M$ trough the coefficient $s_{\alpha}$.}
\begin{align}
\label{eq:w}
\Delta w^{(\alpha)} (M,\lambda) \simeq  \frac{2}{\sqrt{\lambda}} \sum_{k=2}^{\infty}\sum_{\ell=2}^{\infty}\frac{I_k(\sqrt{\lambda})\,I_{\ell}(\sqrt{\lambda})}{I_{1}(\sqrt{\lambda})\,I_{2}(\sqrt{\lambda})}\,\sqrt{k\,\ell}\, (\textsf{D}_{k,\ell}^{(\alpha)}-\delta_{k,\ell})\,.
\end{align}

\begin{figure}
\centering
\begin{tikzpicture}
\draw[black, thick] (0,0) -- (4,0);
\filldraw (2,0) circle (3pt);
\draw (0,0) node[left] {$\frac{\sqrt{k}}{N}\,I_{k}(\sqrt{\lambda})$};
\draw (2,0) node[above] {$\textsf{D}_{k,\ell}^{(\alpha)}$};
\draw (4,0) node[right] {$\frac{\sqrt{\ell}}{N}\,I_{\ell}(\sqrt{\lambda})$};
\end{tikzpicture}
\caption{Graphical representation of the two point function \eqref{2ptquiver}. To each line is associated a factor $\frac{\sqrt{p}}{N}I_{p}(\sqrt{\lambda})$, while to the vertex (black dot) is associated the term $\textsf{D}_{k,\ell}^{(\alpha)}$. The sums over $k$ and $\ell$ are understood.}
\label{fig:2points}
\end{figure}

\subsection{Strong coupling limit}
In order to determine the strong coupling limit of \eqref{eq:w} , we follow the same procedure applied in \cite{Pini:2023svd}. As a first step, we divide the sums over $k$ and $\ell$  considering separately the odd and even contributions. Then we expand the ratios between Bessel functions in \eqref{eq:w} for large values of $\lambda$ using polynomials $Q_{2s}^{(j)\,\text{odd}}(k)$ and $Q_{2s}^{(j)\,\text{even}}(k)$, that read
\begin{align}
\frac{I_{2k+1}(\sqrt{\lambda})}{I_{j}(\sqrt{\lambda})} \equiv \sum_{s=0}^{\infty}\frac{Q^{(j)\,\text{odd}}_{2s}(k)}{\lambda^{s/2}}, \ \ \ \ \ \ \frac{I_{2k}(\sqrt{\lambda})}{I_{j}(\sqrt{\lambda})} \equiv \sum_{s=0}^{\infty}\frac{Q^{(j)\,\text{even}}_{2s}(k)}{\lambda^{s/2}}\,.
\label{IIratios}
\end{align}
Moreover we introduce the functions
\begin{align}
\psi_k(x)=(-1)^{\frac{k}{2}(k-1)}\sqrt{k}\frac{J_k(\sqrt{x})}{\sqrt{x}}  \,,
\end{align}
that, as explained in Appendix \ref{app:Galpha}, allows us to express the non-trivial matrix elements of the $\mathsf{X}$-matrix  \eqref{Xmatrix} as
\begin{align}
\textsf{X}_{k,\ell} = \langle\psi_k |\textsf{X} |\psi_{\ell} \rangle \,,
\end{align}
where $\textsf{X}$ denotes the operator acting on the Hilbert space.\footnote{We refer the reader to \cite{Beccaria:2022ypy} for the definition of this quantity in terms of the Bessel operators 
\begin{equation}
K^{\text{odd}}(x,y) = \sum_{i=1}^{\infty}\psi_{2i+1}(x)\psi_{2i+1}(y), \ \ \ \ K^{\text{even}}(x,y) = \sum_{i=1}^{\infty}\psi_{2i}(x)\psi_{2i}(y) \, \ .   
\end{equation}
} Hence, the strong coupling expansion of the expression \eqref{eq:w} becomes
\begin{align}
&\Delta w^{(\alpha)}(M,\lambda) \, \underset{\lambda \rightarrow \infty}{\sim } \, \frac{2}{\sqrt{\lambda}}\sum_{P=0}^{\infty} \frac{\mathcal{S}^{(P)}(s_{\alpha})}{\lambda^{P/2}} = \frac{2}{\sqrt{\lambda}}\left[S^{(0,0)}(s_{\alpha}) + \frac{S^{(0,1)}(s_{\alpha})+S^{(1,0)}(s_{\alpha})}{\sqrt{\lambda}} \right. \notag \\[0.5em]
& \left. + \frac{S^{(2,0)}(s_{\alpha})+S^{(1,1)}(s_{\alpha})+S^{(0,2)}(s_{\alpha})}{\lambda} + \dots\right]\, ,  
\label{eq:targetodd}
\end{align}
where the coefficients are given by
\begin{align}
\mathcal{S}^{(P)}(s_{\alpha}) = \sum_{L+J=P} S^{(L,J)}(s_{\alpha}) = \sum_{L+J=P} \left( S^{(L,J)}_{\text{odd}}(s_{\alpha})\, +\, S^{(L,J)}_{\text{even}}(s_{\alpha}) \right)
\,,
\end{align}
with
\begin{align}
S_{\text{odd}}^{(L,J)}(s_{\alpha}) & = \sum_{k=1}^{\infty}\sum_{\ell=1}^{\infty}\sqrt{2k+1}\,Q_{2L}^{(1)\,\text{odd}}\,(k)\,\sqrt{2\ell+1}\,Q_{2J}^{(2)\,\text{odd}}(\ell)\langle \psi_{2k+1} \Big| \frac{s_{\alpha}\textsf{X}}{1-s_{\alpha}\textsf{X}} \Big|\psi_{2\ell+1} \rangle \notag \\
& \equiv \langle \Phi^{(1)\,\text{odd}}_L \Big|\frac{s_{\alpha}\textsf{X}}{1-s_{\alpha}\textsf{X}}\Big|\Phi^{(2)\,\text{odd}}_J \rangle\, \ , \\[0.5em]
S_{\text{even}}^{(L,J)}(s_{\alpha}) & = \sum_{k=1}^{\infty}\sum_{\ell=1}^{\infty}\sqrt{2k}\,Q_{2L}^{(1)\,\text{even}}\,(k)\,\sqrt{2\ell}\,Q_{2J}^{(2)\,\text{even}}(\ell)\langle \psi_{2k} \Big| \frac{s_{\alpha}\textsf{X}}{1-s_{\alpha}\textsf{X}} \Big|\psi_{2\ell} \rangle \notag \\
& \equiv \langle \Phi^{(1)\,\text{even}}_L \Big|\frac{s_{\alpha}\textsf{X}}{1-s_{\alpha}\textsf{X}}\Big|\Phi^{(2)\,\text{even}}_J \rangle\,,  
\end{align}
and
\begin{subequations}
\begin{align}
& \Phi^{(j)\,\text{odd}}_L(x) = \sum_{n=1}^{\infty}\sqrt{2n+1}\,Q_{2L}^{(j)\,\text{odd}}(n)\,\psi_{2n+1}(x)\,, \label{Phiodd} \\
& \Phi^{(j)\,\text{even}}_L(x) = \sum_{n=1}^{\infty}\sqrt{2n}\,Q_{2L}^{(j)\,\text{even}}(n)\,\psi_{2n}(x)\,. \label{Phieven} 
\end{align}
\end{subequations}
Based on the expansions for $\Phi^{(1,2)\,\text{odd}}_L(x)$ and $\Phi^{(1,2)\,\text{even}}_L(x)$ collected in Appendix \ref{app:phi}, we find that 
\begin{subequations}
\begin{align}
& \Phi^{(1)\,\text{odd}}_L(x) = [(-2)^{L-1}(x\partial_x)^{L} + (-2)^{L-3}(3+2L-L^2)(x\partial_x)^{L-1} + \dots]\,J_{2}(\sqrt{x})\,, \label{Phi1odd} \\[0.5em]
& \Phi^{(2)\,\text{odd}}_L(x) = [(-2)^{L-1}(x\partial_x)^{L} + (-2)^{L-3}(2L-L^2)(x\partial_x)^{L-1} + \dots]\,J_{2}(\sqrt{x})\,,\\[0.5em]
& \Phi^{(1)\,\text{even}}_L(x) = [(-2)^{L-1}(x\partial_x)^{L} + (-2)^{L-3}(2L-L^2)(x\partial_x)^{L-1} + \dots ]\,J_1(\sqrt{x})\,, \label{Phi1even} \\[0.5em]
&  \Phi^{(2)\,\text{even}}_L(x) = [(-2)^{L-1}(x\partial_x)^{L} + (-2)^{L-3}(-3 +2L -L^2)(x\partial_x)^{L-1} + \dots ]\,J_1(\sqrt{x})\,.
\end{align}
\end{subequations}
At this point it is useful to introduce the coefficients
\begin{align}
\label{wnmalpha}
w^{(\ell)}_{n,m}(s_{\alpha}) \equiv \langle (x\partial_x)^n\phi^{(\ell)}(x) \Big|\frac{s_{\alpha}\textsf{X}}{1-s_{\alpha}\textsf{X}}\Big| (x\partial_x)^{m}\phi^{(\ell)}(x) \rangle \,,
\end{align}
with $n,m \geq 0$ and where $\phi^{(\ell)}(x) \equiv J_\ell(\sqrt{x})$ with $\ell=1,2$. 
These coefficients are the generalization for every $s_{\alpha}$ of the $w_{n,m}$ introduced in \cite{Beccaria:2023kbl} and their main properties have been collected in Appendix \ref{app:Galpha}. Crucially for us, we can express the quantities $\mathcal{S}^{(P)}(s_{\alpha})$ appearing in \eqref{eq:targetodd} using the coefficients \eqref{wnmalpha} and, for instance, for the first values of $P$ we get
\begin{align}
& \mathcal{S}^{(0)}(s_{\alpha}) = S^{(0,0)}(s_{\alpha}) = \frac{1}{4}\, \left(w^{(1)}_{0,0}(s_{\alpha})+w_{0,0}^{(2)}(s_{\alpha})\right)\,, \\[1em]
& \mathcal{S}^{(1)}(s_{\alpha}) = S^{(1,0)}(s_{\alpha})+ S^{(0,1)}(s_{\alpha}) = \frac{1}{8}\,w_{0,0}^{(1)}(s_{\alpha}) -\frac{5}{8}\,w_{0,0}^{(2)}(s_{\alpha}) \notag \\ 
& \ \ \ \ \ \ \ \ \ \ \ \ \ -\frac{1}{2}\left(w^{(1)}_{1,0}(s_{\alpha})+w^{(1)}_{0,1}(s_{\alpha})+w_{1,0}^{(2)}(s_{\alpha})+w_{0,1}^{(2)}(s_{\alpha})\right)\,, \label{S1} \\[1em]
& \mathcal{S}^{(2)}(s_{\alpha})= S^{(2,0)}(s_{\alpha})+ S^{(1,1)}(s_{\alpha})+ S^{(0,2)}(s_{\alpha}) = \frac{35}{32}\,w_{0,0}^{(1)}(s_{\alpha}) -\frac{17}{32}\,w^{(2)}_{0,0}(s_{\alpha}) \notag\\
&  \ \ \ \ \ \ \ \ \ \ \ \ \ -\frac{1}{2}\left(w_{1,0}^{(1)}(s_{\alpha})+w_{0,1}^{(1)}(s_{\alpha})\right)  +(w^{(2)}_{0,1}(s_{\alpha})+w^{(2)}_{1,0}(s_{\alpha}))+(w^{(1)}_{0,2}(s_{\alpha})+w^{(1)}_{1,1}(s_{\alpha}) \notag \\
& \ \ \ \ \ \ \ \ \ \ \ \ \ +w^{(1)}_{2,0}(s_{\alpha}))+(w^{(2)}_{0,2}(s_{\alpha})+w^{(2)}_{1,1}(s_{\alpha})+w^{(2)}_{2,0}(s_{\alpha}))\,, \\[1em]
& \mathcal{S}^{(3)}(s_{\alpha}) = S^{(3,0)}(s_{\alpha}) + S^{(2,1)}(s_{\alpha}) + S^{(1,2)}(s_{\alpha}) + S^{(0,3)}(s_{\alpha}) = -\frac{25}{32}\,w^{(1)}_{0,0}(s_{\alpha})-\frac{1}{32}w^{(2)}_{0,0}(s_{\alpha}) \nonumber \\
& \ \ \ \ \ \ \ \ \ \ \ \ \ +\frac{1}{4}\left(w_{1,0}^{(1)}(s_{\alpha})+w_{0,1}^{(1)}(s_{\alpha})\right)+\frac{5}{2}\left(w^{(2)}_{0,1}(s_{\alpha})+w^{(2)}_{1,0}(s_{\alpha})\right)-\frac{1}{2}(w_{0,2}^{(2)}(s_{\alpha})+w_{2,0}^{(2)}(s_{\alpha}))  \nonumber \\
&\ \ \ \ \ \ \ \ \ \ \ \ \ +\frac{5}{2}\left(w_{0,2}^{(1)}(s_{\alpha})+w_{2,0}^{(1)}(s_{\alpha})\right) -\frac{3}{2}w^{(2)}_{1,1}(s_{\alpha}) +\frac{3}{2}w_{1,1}^{(1)}(s_{\alpha})-2(w_{3,0}^{(1)}(s_{\alpha})+w_{2,1}^{(1)}(s_{\alpha})\notag \\
& \ \ \ \ \ \ \ \ \ \ \ \ \ + w_{1,2}^{(1)}(s_{\alpha})+w_{0,3}^{(1)}(s_{\alpha}))  -2(w_{3,0}^{(2)}(s_{\alpha})+w_{2,1}^{(2)}(s_{\alpha})+w_{1,2}^{(2)}(s_{\alpha})+w_{0,3}^{(2)}(s_{\alpha}))\,.
\end{align}
As shown in Appendix \ref{app:Galpha}, at strong coupling the leading term of the coefficients $w_{n,m}^{(\ell)}(s_{\alpha})$ scales as
\begin{align}
\label{omegalambdabehaviuor}
w_{n,m}^{(\ell)}(s_{\alpha})=\omega_{n,m}^{(\ell)}(s_{\alpha})\left(\frac{\sqrt{\lambda}}{4\pi}\right)^{n+m+1}+\, \ O\bigl((\sqrt{\lambda})^{n+m}\bigr)\,.
\end{align}
Moreover, we observe that the leading order contribution takes the generic form
\begin{align}
(-2)^{P-2}\left(w^{(\ell)}_{P,0}(s_{\alpha}) + w^{(\ell)}_{P-1,1}(s_{\alpha}) + w^{(\ell)}_{P-2,2}(s_{\alpha}) + \dots + w^{(\ell)}_{1,P-1}(s_{\alpha}) + w^{(\ell)}_{0,P}(s_{\alpha})\right)
\end{align}
 for $P=n+m$. Thus, since the leading term of $w^{(\ell)}_{n,m}(s_{\alpha})$ does not depend on $\ell$, henceforth we can omit the dependence on $\ell$ and just write $\omega_{n,m}(s_{\alpha})$. Based on these observations, we are now able to determine the leading order (LO) of the expansion of \eqref{eq:targetodd}. We obtain
\begin{align}
\mathcal{S}^{(P)}_{\text{LO}}(s_{\alpha}) \underset{\lambda\rightarrow\infty}{\sim} 2\sum_{n+m=P}(-2)^{n+m-2}\omega_{n,m}(s_{\alpha})\,(\sqrt{\lambda}/4\pi)^{n+m+1}=\frac{\sqrt{\lambda}}{8\pi}\sum_{n+m=P}\frac{\omega_{n,m}(s_{\alpha})\,(\sqrt{\lambda})^{n+m}}{(-2\pi)^{n}(-2\pi)^{m}}
\end{align}
where the overall factor of 2 is due to the fact that the coefficients $\omega_{n,m}(s_{\alpha})$ are equal for both $S^{(n,m)}_{\text{odd}}(s_{\alpha})$ and $S^{(n,m)}_{\text{even}}(s_{\alpha})$.
Therefore at the leading order we analytically find

\begin{align}
\label{WWZM}
\Delta w^{(\alpha)}(M,\lambda) \underset{\lambda \rightarrow \infty}{\sim} \frac{2}{\sqrt{\lambda}}\sum_{P=0}^{\infty} \frac{\mathcal{S}^{(P)}(s_{\alpha})}{\lambda^{P/2}} & =  \frac{1}{4\pi}\sum_{n=0}^{\infty}\sum_{m=0}^{\infty}\frac{\omega_{n,m}(s_{\alpha})}{(-2\pi)^{n}(-2\pi)^{m}} + O\left(\frac{1}{\sqrt{\lambda}}\right) \notag \\
&\ \ \  = \  \frac{1}{4\pi}G^{(0)}(s_{\alpha},-2\pi,-2\pi) + O\left(\frac{1}{\sqrt{\lambda}}\right) \, .
\end{align}
The generating function $G^{(0)}(s_{\alpha},x,y)$ for the coefficients $\omega_{n,m}(s_{\alpha})$ is defined in Appendix \ref{app:Galpha}, where it is also argued that
\begin{align}
G^{(0)}(s_{\alpha},-2\pi,-2\pi) = \frac{\pi}{s_{\alpha}}\mathcal{I}_0(s_{\alpha})^2-4\pi\, \, ,  
\end{align}
where the integral $\mathcal{I}_0(s_{\alpha})$ has been defined in  \eqref{Inalpha}. 
Therefore we find that
\begin{align}
\label{eq:WWZMfinal}
1+\Delta w^{(\alpha)}(M,\lambda) \, \underset{\lambda\rightarrow\infty}{\sim} \, \frac{1}{s_{\alpha}} \left(\frac{\mathcal{I}_0(s_{\alpha})}{2}\right)^2 + O\left(\frac{1}{\sqrt{\lambda}}\right)\,.
\end{align}
This our final expression for the leading term of the large-$\lambda$ expansion of the ratio \eqref{eq:ratioWW}. For example for the $\mathbb{Z}_2$ quiver gauge theory $s_{\alpha}=1$ and the expression \eqref{eq:WWZMfinal} reads
\begin{align}
\label{eq:WWZ2final}
1+\Delta w^{(1)}(2,\lambda) \, \underset{\lambda\rightarrow\infty}{\sim} \, \frac{\pi^2}{16} +O\left(\frac{1}{\sqrt{\lambda}}\right)\,.
\end{align}

\subsubsection{Numerical checks for the \texorpdfstring{$\mathbb{Z}_2$}{} quiver gauge theory}
\label{subsec:checknumZ2}
\begin{figure}[ht!]
\center{
\includegraphics[scale=0.40]{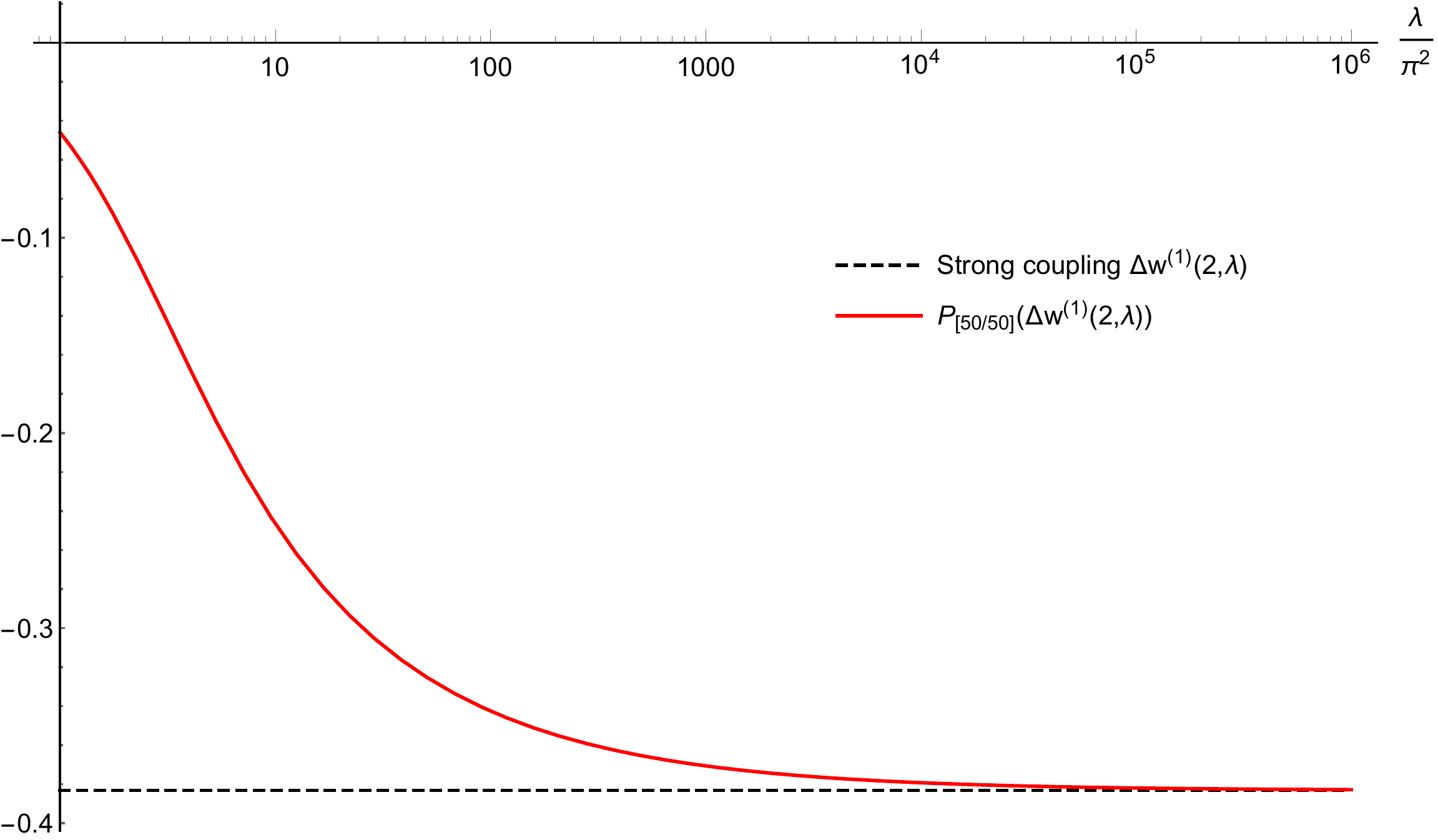}
\caption{Comparison between the large-$\lambda$ theoretical prediction \eqref{eq:WWZ2final} $\Delta w^{(1)}(2,\lambda) = \frac{\pi^2}{16}-1 \simeq -0.38315 $ (black dashed line) and the diagonal Padé curve with $q=50$ for the function $\Delta w^{(1)}(2,\lambda)$ (red line).}
\label{fig:Padecheck}
}
\end{figure}

Here we provide some numerical checks for the $\mathbb{Z}_2$ quiver gauge theory. We follow the same procedure described in Section 4 of \cite{Pini:2023svd}. As a first step, using the perturbative expansion of the elements of the $\textsf{X}$-matrix \eqref{Xmatrix}, we have generated very long series for the coefficients $\textsf{D}_{k,\ell}^{(1)}$ with $k,\ell \geq 2$. For example the first orders of the series expansion for $\textsf{D}_{2,2}^{(1)}$ read
\begin{align}
\label{D22perturbative}
\textsf{D}_{2,2}^{(1)} = 1 -\frac{3\,\zeta_3}{32\pi^4}\lambda^2 + \frac{5\,\zeta_5}{64\pi^6}\lambda^3 + \frac{1}{4096\pi^8}\left(36\,\zeta_3^2-245\,\zeta_7\right)\lambda^4 + \cdots \, \ ,
\end{align}
where the dots stand for higher orders of the expansion. 
Then, using these expressions, we have obtained the perturbative expansion of the $\lambda$-dependent part of \eqref{eq:w}, namely
\begin{align}
\label{eq:SeriesNumZ2}
\Delta w^{(1)}(2,\lambda) = \frac{2}{\sqrt{\lambda}}\sum_{k=0}^{W}\sum_{\ell=0}^{W}\frac{I_k(\sqrt{\lambda})I_{\ell}(\sqrt{\lambda})}{I_1(\sqrt{\lambda})I_2(\sqrt{\lambda})}\sqrt{k\,\ell}\left(\textsf{D}_{k,\ell}^{(1)}-\delta_{k,\ell}\right) \equiv   \frac{2}{\sqrt{\lambda}}\sum_{p=0}^{W}\mathcal{C}_{\mathbb{Z}_2}^{(p)}\left(\frac{\lambda}{\pi^2}\right)^{p} \, ,
\end{align}
where the $\mathcal{C}_{\mathbb{Z}_2}^{(p)}$ are numerical coefficients and the summation stops at some finite cut-off $W$. In turns this implies that we need to compute the expansion of the coefficients $\textsf{D}^{(1)}_{k,\ell}$ up to the order $\lambda^{W}$. We choose to fix $W=110$; this value represents a good compromise among the need to generate enough precise numerical results and the related computational cost. Using the ratio test we observe that the series \eqref{eq:SeriesNumZ2} has a finite radius of convergence located at $\lambda \simeq \pi^2$. Nevertheless we can extend it beyond this limit with a Padé resummation. For this reason we consider the diagonal Padé approximant
\begin{align}
P_{[q/q]}(\Delta w^{(1)}(2,\lambda)) \equiv \left[\sum_{p=0}^{110}\mathcal{C}_{\mathbb{Z}_2}^{(p)}\left(\frac{\lambda}{\pi^2}\right)^p\right]_{[q/q]} \, .
\end{align}
As a last step, we find useful to consider a conformal Padé, that has the advantage to be very stable even for very high values of the coupling $\lambda$ \cite{Beccaria:2021vuc,Costin:2019xql,Costin:2020hwg}. Following \cite{Beccaria:2021hvt} we perform the replacement
\begin{align}
    \frac{\lambda}{\pi^2} \mapsto \frac{4z}{(z-1)^2} \,,
\end{align}
then we construct the Padé approximant in the $z$ variable inside the unit circle $|z| \leq 1$. Finally we express the result as a function of $\lambda$ using the inverse map
\begin{align}
z = \frac{\sqrt{1+\frac{\lambda}{\pi^2}}-1}{\sqrt{1+\frac{\lambda}{\pi^2}}+1} \,  .
\end{align}
The result of this analysis is reported in Figure 
 \ref{fig:Padecheck}. We observe that for very large values of $\lambda$ the Padé curve tends to the constant value predicted by \eqref{eq:WWZ2final}. We regard this numerical result as a strong confirmation of our theoretical strong coupling prediction.

\section{The 3-point function \texorpdfstring{$\langle\,W_{\alpha}\,W_{\beta}\,W_{\alpha+\beta}^{\dagger}\rangle$}{} }
\label{sec: 3pt}
In $\mathbb{Z}_M$ quiver gauge theories with $M>2$ it appears a new irreducible correlator among twisted Wilson loops, namely the 3-point function 
\begin{align}
\langle\, W_{\alpha}\,W_{\beta}\,W^{\dagger}_{\alpha+\beta}\, \rangle \,.
\end{align}
In this section we first  study the main properties of the 3-point correlator of twisted Wilson loops in the planar limit, focusing on its strong coupling regime, and then we will examine the mixed 3-point function with one untwisted and two twisted Wilson loops. 

\begin{figure}
\centering
    \begin{tikzpicture}
     \draw[black, thick] (0,0) -- (0,-2);
     \draw[black, thick] (0,0) -- (-1.41,1.41);
     \draw[black,thick]   (0,0)--(1.41,1.41);
     \draw[black, fill=white] (0,0) circle[radius=4pt];
     \draw (0,-2) node[below] {$\frac{\sqrt{k_3}}{N}\,I_{k_3}(\sqrt{\lambda})$};
    \draw (-1.41,1.41) node[left] {$\frac{\sqrt{k_1}}{N}\,I_{k_1}(\sqrt{\lambda})$};
     \draw (1.41,1.41) node[right] {$\frac{\sqrt{k_2}}{N}\,I_{k_2}(\sqrt{\lambda})$};
    \end{tikzpicture}
    \caption{Graphical representation of one of the  ``building blocks" of the 3-point function \eqref{eq:ZMwawbwcFINAL}. To each line is associated a factor $\frac{\sqrt{k_i}}{N}I_{k_i}(\sqrt{\lambda})$, while to the vertex (white circle) is associated the factor $\textsf{d}_{k_1}^{(\alpha_1)}\textsf{d}_{k_2}^{(\alpha_2)}\textsf{d}_{k_3}^{(\alpha_3)}/(\sqrt{M}N)$.}
    \label{fig:3point}
\end{figure}
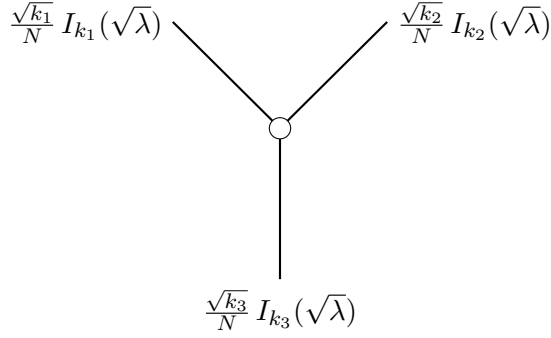

Using \eqref{defWL} and \eqref{Atwisted}, we obtain
\begin{align}
\label{ZMWaWbWc}
\langle W_\alpha\, W_\beta\, W_{\alpha+\beta}^{\dagger} \rangle = \frac{1}{N^3}\sum_{k_1=0}^{\infty}\sum_{k_2=0}^{\infty}\sum_{k_3=0}^{\infty}\frac{1}{k_1!\,k_2!\,k_3!}\left(\frac{\lambda}{2N}\right)^{\frac{k_1+k_2+k_3}{2}}\langle \, \hat{A}_{\alpha,k_1}\,\hat{A}_{\beta,k_2}\,\hat{A}_{\alpha+\beta,k_3}^{\dagger}\, \rangle \, .
\end{align}
We utilise the change of basis \eqref{fromAtoP} and in the planar limit we obtain 
\begin{align}
& \langle \, \hat{A}_{\alpha,k_1}\,\hat{A}_{\beta,k_2}\,\hat{A}_{\alpha+\beta,k_3}^{\dagger}\, \rangle \simeq \left(\frac{N}{2}\right)^{\frac{k_1+k_2+k_3}{2}}\sum_{i=0}^{\lfloor \frac{k_1-1}{2} \rfloor}\sqrt{k_1-2i}\left(\begin{array}{c}
     k_1  \\
     i
\end{array}\right) \sum_{j=0}^{\lfloor \frac{k_2-1}{2} \rfloor}\sqrt{k_2-2j}\left(\begin{array}{c}
     k_2  \\
     j
\end{array}\right) \nonumber \\
& \sum_{\ell=0}^{\lfloor \frac{k_3-1}{2} \rfloor}\sqrt{k_3-2\ell}\left(\begin{array}{c}
     k_3  \\
     \ell
\end{array}\right)\langle \mathcal{P}_{\alpha,k_1-2i}\,\mathcal{P}_{\beta,k_2-2j}\,\mathcal{P}_{\alpha+\beta,k_3-2\ell}^{\dagger} \rangle\, \delta_{\text{Mod}(k_1+k_2+k_3,2),0} \,,
\end{align}
where the Kronecker delta function is needed to enforce that, as discussed in \cite{Billo:2022fnb},  $k_1+k_2+k_3$ must be even. Then we use \eqref{3pointP} 
and we write
 \begin{align}
\langle \mathcal{P}_{\alpha,k_1-2i}\,\mathcal{P}_{\beta,k_2-2j}\,\mathcal{P}_{\alpha+\beta,k_3-2\ell}^{\dagger} \rangle \simeq \frac{1}{\sqrt{M}}\frac{1}{N}\,\textsf{d}^{(\alpha)}_{k_1-2i}\,\textsf{d}_{k_2-2j}^{(\beta)}\,\textsf{d}_{k_3-2\ell}^{(\alpha+\beta)}\,,
 \end{align}
where $\textsf{d}^{(\alpha)}_k$ was defined in \eqref{dalfa} 
This way the planar limit of the correlator \eqref{ZMWaWbWc} factorizes in the product of three contributions, namely
\begin{align}
\label{WaWbWa+bplanar}
& \langle W_\alpha\, W_\beta\, W_{\alpha+\beta}^{\dagger}\rangle \simeq \nonumber \\
& \frac{1}{\sqrt{M}\,N^4}\prod_{p=1}^{3}\left[\sum_{k_p=0}^{\infty}\frac{1}{k_p!}\left(\frac{\lambda}{2N}\right)^{\frac{k_p}{2}}\left(\frac{N}{2}\right)^{\frac{k_p}{2}}\sum_{i=0}^{\lfloor \frac{k_p-1}{2} \rfloor}\sqrt{k_p-2i}\left(\begin{array}{c}
     k_p  \\
     i 
\end{array}\right)
\textsf{d}^{(\alpha_p)}_{k_p-2i}\right]\delta_{\text{Mod}(k_1+k_2+k_3,2),0} \,,
\end{align}
where we introduced
\begin{align}
\label{eq:defabc}
\alpha_1 \equiv \alpha, \ \ \ \ \ \alpha_2 \equiv \beta, \ \ \ \ \  \alpha_3 \equiv M-\alpha-\beta\,. 
\end{align}
Now we perform the sums over $k_p$. We notice that it is convenient to treat separately the case of even $k_p$ and odd $k_p$. For $k_p=2k$  we have
\begin{align}
&\sum_{k=0}^{\infty}\frac{1}{(2k)!}\left(\frac{\sqrt{\lambda}}{2}\right)^{2k}\sum_{i=0}^{k-1}\sqrt{2k-2i}\left(\begin{array}{c}
     2k  \\
     i 
\end{array}\right)
\sum_{\ell=2}^{\infty}\sqrt{\ell}\,\textsf{D}^{(\alpha)}_{2k-2i,\ell}= \, \nonumber \\[0.5em]
&  \sum_{j=2}^{\infty}\sum_{\ell=2}^{\infty}I_{j}(\sqrt{\lambda})\sqrt{j\,\ell}\,\textsf{D}^{(\alpha)}_{j,\ell} =  \sum_{k=1}^{\infty}\sum_{\ell=1}^{\infty}I_{2k}(\sqrt{\lambda})\sqrt{(2k)\,(2\ell)}\,\textsf{D}^{(\alpha)}_{2k,2\ell}   \, \ ,
\label{evenfZM}
\end{align}
where in the second line we used the properties of the Bessel functions collected in Appendix \ref{appendix:A}.
Similarly, for $k_p=2k+1$, we obtain
\begin{align}
& \sum_{k=1}^{\infty}\frac{1}{(2k+1)!}\left(\frac{\sqrt{\lambda}}{2}\right)^{2k+1}\sum_{i=0}^{k}\sqrt{2k+1-2i}\left(\begin{array}{c}
     2k+1  \\
     i 
\end{array}\right)
\sum_{\ell=3}^{\infty}\,\sqrt{\ell}\,\textsf{D}^{(\alpha)}_{2k+1-2i,\ell} \, \nonumber = \\[0.5em]
& \sum_{j=3}^{\infty}\sum_{\ell=3}^{\infty}I_{j}(\sqrt{\lambda})\sqrt{(j)\,(\ell)}\,\textsf{D}^{(\alpha)}_{j,\ell} =  \sum_{k=1}^{\infty}\sum_{\ell=1}^{\infty}I_{2k+1}(\sqrt{\lambda})\sqrt{(2k+1)\,(2\ell+1)}\,\textsf{D}^{(\alpha)}_{2k+1,2\ell+1} \,.
\label{oddfZM}
\end{align}   
Finally using both \eqref{evenfZM} and \eqref{oddfZM} we get our final expression for the planar limit of the correlator \eqref{ZMWaWbWc} 
\begin{align}
&  \langle W_\alpha\, W_\beta \, W_{\alpha+\beta}^{\dagger} \rangle \simeq  \frac{1}{\sqrt{M}N^4}\left[ \, \prod_{p=1}^{3}\left( \sum_{k=1}^{\infty}\sum_{\ell=1}^{\infty}I_{2k}(\sqrt{\lambda})\sqrt{(2k)\,(2\ell)}\,\textsf{D}^{(\alpha_p)}_{2k,2\ell}\right)  + \right. \nonumber \\[0.5em]
& \left. \sum_{\sigma \in \mathcal{Q}_3}\left[\left(\sum_{k=1}^{\infty}\sum_{\ell=1}^{\infty}I_{2k}(\sqrt{\lambda})\sqrt{(2k)\,(2\ell)}\,\textsf{D}^{(\alpha_{\sigma(1)})}_{2k,2\ell}\right) \right. \right.  \left(  \sum_{k=1}^{\infty}\sum_{\ell=1}^{\infty}I_{2k+1}(\sqrt{\lambda})\sqrt{(2k+1)\,(2\ell+1)}\,\textsf{D}^{(\alpha_{\sigma(2)})}_{2k+1,2\ell+1} \right) \nonumber \\[0.5em]
& \left. \left. \left(  \sum_{k=1}^{\infty}\sum_{\ell=1}^{\infty}I_{2k+1}(\sqrt{\lambda})\sqrt{(2k+1)\,(2\ell+1)}\,\textsf{D}^{(\alpha_{\sigma(3)})}_{2k+1,2\ell+1} \right) \right]\,  \right] \,,
\label{eq:ZMwawbwcFINAL}
\end{align}
where the set of permutations reads
\begin{align}
\label{setQ3}
\mathcal{Q}_3 = \{\,(1,2,3),\,(3,1,2),\,(2,3,1)\, \}\,.
\end{align}
The structure of the correlator \eqref{eq:ZMwawbwcFINAL} can be efficiently summarized using the diagram reported in Figure \ref{fig:3point}.

Also in this case it is useful to understand which is the corresponding observable in $\mathcal{N}=4$ SYM. For this reason we turn off the interaction action and we set $M=1$, then the expression \eqref{eq:ZMwawbwcFINAL} becomes 
\begin{align}
\frac{\lambda^{3/2}}{8\,N^4}I_{1}(\sqrt{\lambda})\left(I_1(\sqrt{\lambda})^2+3\,I_2(\sqrt{\lambda})^2\right)\, ,
\end{align}
that corresponds to the planar limit of the $\mathcal{N}=4$ connected Wilson loop
\begin{align}
\label{eq:Z3connZM}
W^{(3)}_{conn}(\lambda) \equiv  \langle \,W\,W\,W\, \rangle_{(0)} -3\,\langle\,W \rangle_{(0)}\langle W\,W \rangle_{(0)} + 2\,\langle W\rangle_{(0)}^3  \, .
\end{align}
Therefore we consider the ratio between \eqref{eq:ZMwawbwcFINAL} and \eqref{eq:Z3connZM}, namely
\begin{align}
\label{ratioWWW}
\frac{\langle W_\alpha\,W_\beta\,W_{\alpha+\beta}^{\dagger} \rangle}{\sqrt{M}\,W_{conn}^{(3)}(\lambda)} \equiv 1+\Delta w^{(\alpha,\beta)}(M,\lambda)  \,, 
\end{align}
which is equal to 1 for $S_{\mathrm{int}}\rightarrow 0$ and all the dependence on the coupling is encoded in the function $\Delta w^{(\alpha,\beta)}(M,\lambda)$. We find convenient to divide and multiply the r.h.s. of the above expression  by $I_1(\sqrt{\lambda})^2$. Therefore, in the following, we evaluate the strong coupling behaviour of the expression
\begin{align}
& 1+\Delta w^{(\alpha,\beta)}(M,\lambda) \simeq   \frac{8}{\lambda^{3/2}}\left(\frac{I_1(\sqrt{\lambda})^2}{I_1(\sqrt{\lambda})^2+3I_2(\sqrt{\lambda})^2}\right)\left[\, \prod_{p=1}^{3}\left( \sum_{k=1}^{\infty}\sum_{\ell=1}^{\infty}\frac{I_{2k}(\sqrt{\lambda})}{I_1(\sqrt{\lambda})}\sqrt{4k\,\ell}\,\textsf{D}^{(\alpha_p)}_{2k,2\ell}\right) + \nonumber \right. \nonumber \\[0.5em]
& \left.  \sum_{\sigma \in \mathcal{Q}_3}\left[\left(\sum_{k=1}^{\infty}\sum_{\ell=1}^{\infty}\frac{I_{2k}(\sqrt{\lambda})}{I_1(\sqrt{\lambda})}\sqrt{4k\,\ell}\,\textsf{D}^{(\alpha_{\sigma(1)})}_{2k,2\ell}\right)\left(  \sum_{k=1}^{\infty}\sum_{\ell=1}^{\infty}\frac{I_{2k+1}(\sqrt{\lambda})}{I_1(\sqrt{\lambda})}\sqrt{(2k+1)\,(2\ell+1)}\,\textsf{D}^{(\alpha_{\sigma(2)})}_{2k+1,2\ell+1} \right) \right. \right. \nonumber \\[0.5em]
& \left. \left. \left(  \sum_{k=1}^{\infty}\sum_{\ell=1}^{\infty}\frac{I_{2k+1}(\sqrt{\lambda})}{I_1(\sqrt{\lambda})}\sqrt{(2k+1)\,(2\ell+1)}\,\textsf{D}^{(\alpha_{\sigma(3)})}_{2k+1,2\ell+1}) \right)\right]
\right] \,.
\label{ratioWaWbWc}
\end{align}

\subsection{Strong coupling limit}
In order to analyse the strong coupling limit of  the ratio \eqref{ratioWaWbWc}, we use the large-$\lambda$ expansions \eqref{IIratios} between ratios of Bessel functions and we  introduce the following quantities 
\begin{subequations}
\begin{align}
& \sum_{k=1}^{\infty}\sum_{\ell=1}^{\infty}\frac{I_{2k}(\sqrt{\lambda})}{I_1(\sqrt{\lambda})}\sqrt{(2k)(2\ell)}(\textsf{D}_{2k,2\ell}^{(\alpha)}-\delta_{2k,2\ell}) \equiv \sum_{P=0}^{\infty} \frac{\mathcal{S}_{\text{even}}^{(P)}(s_{\alpha})}{\lambda^{P/2}}\,, \label{expEVEN}\\
&  \sum_{k=1}^{\infty}\sum_{\ell=1}^{\infty}\frac{I_{2k+1}(\sqrt{\lambda})}{I_1(\sqrt{\lambda})}\sqrt{(2k+1)(2\ell+1)}(\textsf{D}_{2k+1,2\ell+1}^{(\alpha)}-\delta_{2k+1,2\ell+1}) \equiv \sum_{P=0}^{\infty} \frac{\mathcal{S}_{\text{odd}}^{(P)}(s_{\alpha})}{\lambda^{P/2}}\,, \label{expODD}
\end{align}
\end{subequations}
where
\begin{align}
& \mathcal{S}_{\text{even}}^{(P)}(s_{\alpha}) =\notag \\
& \sum_{k=1}^{\infty}\sum_{\ell=1}^{\infty}\sqrt{(2k)(2\ell)}\,Q_{2P}^{(1)\,\text{even}}(k)\, \langle \psi_{2k} \Big| \frac{s_{\alpha}\,\textsf{X}}{1-s_{\alpha}\,\textsf{X}} \Big| \psi_{2\ell} \rangle \equiv -\frac{1}{2}\langle \Phi_P^{(1)\,\text{even}} \Big| \frac{s_{\alpha}\textsf{X}}{1-s_{\alpha}\textsf{X}} \Big| \phi^{(1)}(x) \rangle\,, \label{SevenZM} \\[1em]
& \mathcal{S}_{\text{odd}}^{(P)}(s_{\alpha}) = \notag \\
& \sum_{k=1}^{\infty}\sum_{\ell=1}^{\infty}\sqrt{(2k+1)(2\ell+1)}\,Q_{2P}^{(1)\,\text{odd}}(k)\, \langle \psi_{2k+1} \Big| \frac{s_{\alpha}\,\textsf{X}}{1-s_{\alpha}\,\textsf{X}} \Big| \psi_{2\ell+1} \rangle \equiv  -\frac{1}{2}\langle \Phi_P^{(1)\,\text{odd}} \Big| \frac{s_{\alpha}\textsf{X}}{1-s_{\alpha}\textsf{X}} \Big| \phi^{(2)}(x) \rangle \,, \label{SoddZM}
\end{align}
where $\Phi^{(1) \, \text{odd}}_{P}(x)$ and $\Phi^{(1) \, \text{even}}_{P}(x)$ have been defined by the expressions \eqref{Phiodd}-\eqref{Phieven} and $\phi^{(1,2)}(x) \equiv J_{1,2}(\sqrt{x})$. We can express both \eqref{SevenZM} and \eqref{SoddZM} in terms of the  coefficients \eqref{wnmalpha}, 
this way, using the relations \eqref{Phi1odd}, \eqref{Phi1even} and \eqref{omegalambdabehaviuor}, we determine the leading order (LO) of both $\mathcal{S}^{(P)}_{\text{even}}(s_{\alpha})$ and $\mathcal{S}^{(P)}_{\text{odd}}(s_{\alpha})$, i.e. 
\begin{align}
\mathcal{S}^{(P)}_{\text{even}}(s_{\alpha}) \big |_{\text{LO}} = (-2)^{P-2}\,\omega_{P,0}(s_{\alpha})\left(\frac{\sqrt{\lambda}}{4\pi}\right)^{P+1},\ \ \ \ \mathcal{S}^{(P)}_{\text{odd}}\big |_{\text{LO}}(s_{\alpha}) = (-2)^{P-2}\,\omega_{P,0}(s_{\alpha})\left(\frac{\sqrt{\lambda}}{4\pi}\right)^{P+1}\,.
\end{align}
Then, we express the leading term of the quantities \eqref{expEVEN} and \eqref{expODD} in terms of the generating function $G^{(0)}(s_{\alpha},x)$, namely
\begin{align}
\sum_{P=0}^{\infty}\frac{\mathcal{S}_{\text{even}}^{(P)}(s_{\alpha})}{\lambda^{P/2}} = {16\pi}\,G^{(0)}\left(s_{\alpha},-2\pi\right) + O(\lambda^{0}) \, , \ \ \ \   \sum_{P=0}^{\infty}\frac{\mathcal{S}_{\text{odd}}^{(P)}(s_{\alpha})}{\lambda^{P/2}} ={16\pi}\,G^{(0)}\left(s_{\alpha},-2\pi\right)  + O(\lambda^0) \,.
\end{align}
In appendix B we argue that
\begin{align}
\label{G02pi}
G^{(0)}(s_{\alpha},-2\pi) = -\frac{4\pi}{\sqrt{s_{\alpha}}}\left(\mathcal{I}_0(s_{\alpha})+2\sqrt{s_{\alpha}}\right)\, .
\end{align}
This way we determine the leading term of the strong coupling expansion of  \eqref{ratioWWW}, which reads
\begin{align}
& 1+\Delta w^{(\alpha,\beta)}(M,\lambda) \, \underset{\lambda\rightarrow\infty}{\sim} \, -\frac{1}{8}\prod_{p=1}^{3}\frac{\mathcal{I}_0(s_{\alpha_p})}{\sqrt{s_{\alpha_p}}} + O\left(\frac{1}{\sqrt{\lambda}}\right) \, ,
\label{WWWabcSTRONG}
\end{align}
where $\alpha_p$ are defined in \eqref{eq:defabc}.
This is our final expression for the strong coupling limit of the ratio \eqref{ratioWWW}. For example for the $\mathbb{Z}_3$ quiver gauge theory all the $s_{\alpha_{p}}=3/4$ and the 
expression \eqref{WWWabcSTRONG} reads
\begin{align}
1+\Delta w^{(1,1)}(3,\lambda) \underset{\lambda\rightarrow\infty}{\sim} \left(\frac{4\pi}{9\sqrt{3}}\right)^3 +O\left(\frac{1}{\sqrt{\lambda}}\right)\,.
\end{align}

\subsection{Mixed 3-point correlator}
\label{mixed3pt}
There is another non-trivial 3-point function which has to be considered, namely the correlator with one untwisted Wilson loop and two Wilson loops in conjugated twisted sectors. Hence we have
\begin{align}
\label{ZMW0WbWc}
\langle W_0\, W_\alpha\, W_{\alpha}^{\dagger} \rangle = \frac{1}{N^3}\sum_{k_1=0}^{\infty}\sum_{k_2=0}^{\infty}\sum_{k_3=0}^{\infty}\frac{1}{k_1!\,k_2!\,k_3!}\left(\frac{\lambda}{2N}\right)^{\frac{k_1+k_2+k_3}{2}}\langle \, A_{0,k_1}\,\hat{A}_{\alpha,k_2}\,\hat{A}_{\alpha,k_3}^{\dagger}\, \rangle \,.
\end{align}
Using \eqref{Ahatoperators}, this expression is given by the sum of two contributions
\begin{align}
& \langle W_0\, W_\alpha\, W_{\alpha}^{\dagger} \rangle  = \notag \\
& \frac{1}{N^3}\sum_{k_1=0}^{\infty}\sum_{k_2=0}^{\infty}\sum_{k_3=0}^{\infty}\frac{1}{k_1!\,k_2!\,k_3!}\left(\frac{\lambda}{2N}\right)^{\frac{k_1+k_2+k_3}{2}}\left(\langle \, \hat{A}_{0,k_1}\,\hat{A}_{\alpha,k_2}\,\hat{A}_{\alpha,k_3}^{\dagger}\, \rangle +\langle \, A_{0,k_1}\rangle_0 \,\langle \hat{A}_{\alpha,k_2}\,\hat{A}_{\alpha,k_3}^{\dagger}\, \rangle \right) \,.
\label{ZMW0WbWc2}
\end{align}
Let us focus on the second contribution on the r.h.s. of \eqref{ZMW0WbWc2}. We set $k_{1}=2\ell$ (otherwise $\langle A_{0,k_1} \rangle_0$ is vanishing) so that we get
\begin{align}
\label{secondterm1}
& \frac{\sqrt{M}}{N^3}\sum_{\ell=0}^{\infty}\sum_{k_2=0}^{\infty}\sum_{k_3=0}^{\infty}\frac{1}{(2\ell)!\,k_2!\,k_3!}\left(\frac{\lambda}{2N}\right)^{\ell+\frac{k_2+k_3}{2}} t_{2\ell} \,\langle \hat{A}_{\alpha,k_2}\,\hat{A}_{\alpha,k_3}^{\dagger}\, \rangle \,, 
\end{align}
then, using \eqref{2ptquiver}, the expression \eqref{secondterm1} becomes in the planar limit 
\begin{align}
\label{secondterm}
\sqrt{M}\,\langle W \rangle_0\,\frac{1}{N^2}\sum_{k_{2}=2}^{\infty}\sum_{k_{3}=2}^{\infty}I_{k_{2}}(\sqrt{\lambda})I_{k_{3}}(\sqrt{\lambda})\sqrt{k_{2}\,k_{3}}\,\textsf{D}^{(\alpha)}_{k_{2},k_{3}}\,.
\end{align}
Finally we observe that for the first contribution on the r.h.s of \eqref{ZMW0WbWc2} we can perform the same steps as in the previous subsection, which lead to show that in the planar limit this term scales as $O\bigl(\frac{1}{N^4} \bigr)$ (see \eqref{WaWbWa+bplanar}). 
Therefore the term in \eqref{secondterm} is leading with respect to the other in the planar limit and we conclude that
\begin{align}
\langle W_0\, W_\alpha\, W_{\alpha}^{\dagger} \rangle \simeq \sqrt{M}\,\langle W \rangle_0\,\frac{1}{N^2}\sum_{k_{2}=2}^{\infty}\sum_{k_{3}=2}^{\infty}I_{k_{2}}(\sqrt{\lambda})I_{k_{3}}(\sqrt{\lambda})\sqrt{k_{2}\,k_{3}}\,\textsf{D}^{(\alpha)}_{k_{2},k_{3}}\simeq \sqrt{M}\, \langle W \rangle_0\,\langle W_{\alpha}W_{\alpha}^{\dagger}\rangle \,,
\end{align}
which means that the study of  $\langle W_0W_{\alpha}W_{\alpha}^{\dagger} \rangle$ is reduced to the 2-point function that we analysed in Section \ref{sec:2pt}.

\section{Higher point correlators}
\label{sec: npoint}

In this section we perform the computation of higher point correlators. Specifically, we first deal with the case of correlation functions involving only twisted Wilson loops, secondly we conclude with the analysis of mixed correlators, i.e. correlators of both untwisted and twisted Wilson loops. As a warm-up example, we first consider the $\mathbb{Z}_2$ quiver and then we generalize our analysis to the $\mathbb{Z}_M$ quiver.

\subsection{Higher correlators in the \texorpdfstring{$\mathbb{Z}_2$}{} quiver gauge theory}
We start considering the case of observables involving only twisted Wilson loops in the $\mathbb{Z}_2$ quiver gauge theory. We observe that a correlator among an odd number of twisted Wilson loops vanishes. On the other hand, in order to evaluate a correlator among an even number of twisted Wilson loops, we have to consider the following correlation function
\begin{align}
\label{Aeven}
    \langle \, \mathcal{P}_{1,k_1}\,\mathcal{P}_{1,k_2} \dots \mathcal{P}_{1,k_{2n}}\, \rangle \,.
\end{align}
As shown in \cite{Billo:2022fnb} and reviewed in Section \ref{matrixmodel}, in the planar limit correlators of the form \eqref{Aeven} are in general reducible and can be rewritten in terms of the product of 2-point correlators. 
In some cases the coefficient of the leading term of the correlator \eqref{Aeven} vanishes, thus the corresponding correlator is no longer reducible and it is subleading with respect to the reducible ones. Since we just focus on  the planar limit, in the following we only consider contributions arising from reducible Wilson loop correlators. Before addressing the general case, as a pedagogical example, we first examine with some details the case of the 4-point function.

\subsubsection{The 4-point function \texorpdfstring{$\langle W_1\,W_1\,W_1\,W_1 \rangle$}{}}

We want to compute
\begin{align}
\label{4W1}
& \langle \, W_1\,W_1\,W_1\,W_1\, \rangle = \frac{1}{N^4}\sum_{k_1=2}^{\infty}\sum_{k_2=1}^{\infty}\sum_{k_3=1}^{\infty}\sum_{k_4=1}^{\infty}\frac{1}{k_1!\,k_2!\,k_3!\,k_4!\,}\left(\frac{\lambda}{2N}\right)^{\frac{k_1+k_2+k_3+k_4}{2}}\, \langle\, \hat{A}_{1,k_1}\,\hat{A}_{1,k_2}\,\hat{A}_{1,k_3}\,\hat{A}_{1,k_4}\, \rangle  \nonumber \\
& \simeq 3\left(\frac{1}{N^2}\sum_{k_1=2}^{\infty}\sum_{k_2=2}^{\infty}\,I_{k_1}(\sqrt{\lambda})\,I_{k_2}(\sqrt{\lambda})\,\sqrt{k_1\,k_2}\,\textsf{D}_{k_1,k_2}^{(1)}\right)^2 = 3\,\langle W_1\,W_1 \rangle\, \langle \,W_1\,W_1\, \rangle \,,
\end{align}
where we employed the change of basis \eqref{fromAtoP} and we used the decomposition of the 4-point correlator in the planar limit as in \eqref{Example4point}. As in the previous sections, we consider the $\mathcal{N}=4$ observable that is obtained from \eqref{4W1} turning off the interaction action $S_{\mathrm{int}}$, namely  
\begin{align}
\label{W4}
W^{(4)}(\lambda) \equiv \frac{1}{N^4}\sum_{k_1=0}^{\infty}\sum_{k_2=0}^{\infty}\sum_{k_3=0}^{\infty}\sum_{k_4=0}^{\infty}\frac{1}{k_1!\,k_2!\,k_3!\,k_4!}\left(\frac{\lambda}{2N}\right)^{\frac{k_1+k_2+k_3+k_4}{2}}\langle A_{1,k_1}\,A_{1,k_2}\,A_{1,k_3}\,A_{1,k_4} \rangle_0 \,.
\end{align}
Differently from the $\langle W_1\,W_1 \rangle$ case, we observe that the \eqref{W4} does not correspond to an $\mathcal{N}=4$ connected correlator. Nevertheless, using  the relation \eqref{4ptA}, we obtain
\begin{align}
W^{(4)}(\lambda) \simeq 3 \,(W_{conn}^{(2)}(\lambda))^2\,,     
\end{align}
where $W_{conn}^{(2)}(\lambda)$ is given by \eqref{Wconnected}.
Finally we consider the ratio between $\langle W_1\,W_1\,W_1\,W_1 \rangle$ and $W^{(4)}(\lambda)$, which reads
\begin{align}
\label{ratio4}
\frac{\langle\, W_1\,W_1\,W_1\,W_1 \,\rangle}{W^{(4)}(\lambda)} \simeq \left(\frac{\langle W_1\,W_1 \rangle}{W^{(2)}_{conn}(\lambda)}\right)^2 \,.
\end{align}
The expression \eqref{ratio4} can be evaluated at strong coupling using \eqref{eq:WWZ2final}. We find
\begin{align}
 \frac{\langle\, W_1\,W_1\,W_1\,W_1 \,\rangle}{W^{(4)}(\lambda)} \,   \underset{\lambda\rightarrow\infty}{\sim}  \, \left(\frac{\pi}{4}\right)^4 + O\left(\frac{1}{\sqrt{\lambda}}\right) \,.
\end{align}

\subsubsection{The general case \texorpdfstring{$\langle \, W_1\,W_1\, \dots\, W_1\, \rangle $}{}}
The study of the $n$-point correlator can be performed considering the ratio between a correlator involving $2n$ twisted Wilson loops and the $\mathcal{N}=4$ observable $W^{(2n)}(\lambda) \simeq (W^{(2)}_{conn}(\lambda))^n$ obtained turning off the interaction action, i.e.
\begin{align}
\label{eq:Z2general}
\frac{\langle W_1\,W_1\,\dots W_1 \rangle}{W^{(2n)}(\lambda)} \simeq \left(\frac{\langle W_1\,W_1 \rangle}{W^{(2)}_{conn}(\lambda)}\right)^{n} \,.
\end{align}
In the strong coupling limit, exploiting \eqref{eq:WWZ2final}, we easily get the result

\begin{align}
 \frac{\langle W_1\,W_1\,\dots W_1 \rangle}{W^{(2n)}(\lambda)}  \,\underset{\lambda\rightarrow\infty}{\sim} \, \left(\frac{\pi}{4}\right)^{2n} + O\left(\frac{1}{\sqrt{\lambda}}\right) \,.
\end{align}

\subsection{Higher correlators in the  \texorpdfstring{$\mathbb{Z}_M$}{} quiver gauge theory}
We extend the results obtained for the $\mathbb{Z}_2$ quiver gauge theory to the most general $\mathbb{Z}_M$ quiver. Unlike before, now we have to consider also correlators among an odd number of twisted Wilson loops. Since the derivation is slightly different, we split the discussion for correlators with even and odd number of twisted Wilson loops. 

\subsubsection{Even reducible correlators}
Let us start with the case of a correlator among $2n$ Wilson loops, namely
\begin{align}
\label{W2n}
\langle\, W_{\alpha_1}\,W_{\alpha_2}\dots W_{\alpha_{2n}}^{\dagger} \,\rangle \, = \frac{1}{N^{2n}}\sum_{k_1=0}^{\infty}\dots\sum_{k_{2n}=0}^{\infty} \left[\prod_{i=1}^{2n}\frac{1}{k_i!}\left(\frac{\lambda}{2N}\right)^{\frac{k_i}{2}}\right]\,\langle A_{\alpha_1,k_1}\,A_{\alpha_2,k_2}\dots A_{\alpha_{2n},k_{2n}}^{\dagger} \rangle \,.
\end{align}
We use the \eqref{fromAtoP} to expand the correlator \eqref{W2n} on the $\mathcal{P}_{\alpha_i,k_i}$ basis. In the large-$N$ limit it holds that
\begin{align}
& \langle\, \mathcal{P}_{\alpha_1,k_1}\,\mathcal{P}_{\alpha_2,k_2}\dots \mathcal{P}_{\alpha_{2n-1},k_{2n-1}}\,\mathcal{P}_{\alpha_{2n},k_{2n}}^{\dagger} \rangle \simeq \nonumber \\[0.5em]
& \left(\delta_{\alpha_1,\alpha_2}\,\textsf{D}_{k_1,k_2}^{(\alpha_1)}\right)\left(\delta_{\alpha_3,\alpha_4}\,\textsf{D}_{k_3,k_4}^{(\alpha_3)}\right) \dots  \left(\delta_{\alpha_{2n-1},\alpha_{2n}}\,\textsf{D}_{k_{2n-1},k_{2n}}^{(\alpha_{2n-1})}\right) \, +  \, \text{other Wick contractions}\,.
\end{align}
We perform the sums over $k_{i}$ and the expression \eqref{W2n} in the planar limit becomes
\begin{align}
& \langle\, W_{\alpha_1}\,W_{\alpha_2}\dots W_{\alpha_{2n-1}}W_{\alpha_{2n}}^{\dagger} \,\rangle \simeq \nonumber \\
& \prod_{j=1}^{n}\left(\frac{\delta_{\alpha_{2j-1},\alpha_{2j}}}{N^2}\sum_{q=2}^{\infty}\sum_{p=2}^{\infty}\,I_{q}(\sqrt{\lambda})I_{p}(\sqrt{\lambda})\,\sqrt{q\,p}\,\textsf{D}_{q,p}^{(\alpha_{2j-1})}\right)  + \text{other Wick contractions}\,.
\label{eq:W2nLargeN}
\end{align}
We observe that if we turn off the interaction action the expression \eqref{eq:W2nLargeN} gets
\begin{align}
\mathcal{N}^{even}(\alpha_1,\alpha_2,\dots,\alpha_{2n})\left(\frac{\sqrt{\lambda}}{2\,N^2}\right)^n\left[I_1(\sqrt{\lambda})I_2(\sqrt{\lambda})\right]^n \,,
\end{align}
where we used the identity \eqref{idIk} and the overall numerical factor is given by 
\begin{align}
\mathcal{N}^{even}(\alpha_1,\alpha_2,\dots,\alpha_{2n}) \equiv \left(\delta_{\alpha_1,\alpha_2}\delta_{\alpha_3,\alpha_4} \dots \delta_{\alpha_{2n-1},\alpha_{2n}}\,  + \, \text{other Wick contractions}\right) 
\end{align}
and it holds $1 \leq \mathcal{N}^{even}(\alpha_1,\alpha_2,\dots,\alpha_{2n}) \leq (2n-1)!!$\,.
This leads us to define the ratio 
\begin{align}
\label{ratioEVEN}
\frac{\langle\, W_{\alpha_1}\,W_{\alpha_2} \dots W_{\alpha_{2n-1}}W_{\alpha_{2n}}^{\dagger} \,\rangle}{\mathcal{N}^{even}(\alpha_1,\alpha_2,\dots,\alpha_{2n})\,\left(W_{conn}^{(2)}(\lambda)\right)^n} \equiv 
1+\Delta w^{(\alpha_1,\dots,\alpha_{2n-1})} (M,\lambda)  \,,
\end{align}
where $W_{conn}^{(2)}(\lambda)$ was defined in \eqref{Wconnected}. By construction the ratio above is equal to $1$ when we turn off the interaction action.
Then we evaluate the leading contribution of the ratio \eqref{ratioEVEN} in the large-$\lambda$ expansion and we analytically find
\begin{align}
\label{evenFinal}
& 1+\Delta w^{(\alpha_1,\dots,\alpha_{2n-1})} (M,\lambda) \,\underset{\lambda\rightarrow\infty}{\sim} \,  \nonumber \\
& \frac{1}{\mathcal{N}^{even}(\alpha_1,\alpha_2,\dots,\alpha_{2n})}\left[ \prod_{j=1}^{n}\left(\delta_{\alpha_{2j-1},\alpha_{2j}}\frac{\mathcal{I}_0(s_{\alpha_{2j-1}})^2}{4\,s_{\alpha_{2j-1}}}\right) + \text{other Wick contractions} \right] + O\left(\frac{1}{\sqrt{\lambda}}\right)
\end{align} 
where we used the large-$\lambda$ behaviour of the 2-point correlator \eqref{eq:WWZMfinal}. For example for $n=2$ the expression \eqref{evenFinal} reads
\begin{align}
& 1+\Delta w^{(\alpha_1,\alpha_2,\alpha_3)}(M,\lambda) \,\underset{\lambda\rightarrow\infty}{\sim} \, \frac{1}{\mathcal{N}^{even}(\alpha_1,\alpha_2,\alpha_3,\alpha_4)}\left[\delta_{\alpha_1,\alpha_2}\delta_{\alpha_3,\alpha_4}\frac{\mathcal{I}_0(s_{\alpha_1})^2\mathcal{I}_0(s_{\alpha_3})^2}{16\,s_{\alpha_1}s_{\alpha_3}} \right. \nonumber \\[0.5em]
& \left. + \, 
\delta_{\alpha_1,\alpha_3}\delta_{\alpha_2,\alpha_4}\frac{\mathcal{I}_0(s_{\alpha_1})^2\mathcal{I}_0(s_{\alpha_2})^2}{16\,s_{\alpha_1}s_{\alpha_2}} + \delta_{\alpha_1,\alpha_4}\delta_{\alpha_2,\alpha_3}\frac{\mathcal{I}_0(s_{\alpha_1})^2\mathcal{I}_0(s_{\alpha_2})^2}{16\,s_{\alpha_1}s_{\alpha_2}}\right] + O\left(\frac{1}{\sqrt{\lambda}}\right)\,, 
\end{align}
where $\alpha_1,\alpha_2,\alpha_3=1,\dots,M-1$.

\subsubsection{Odd reducible correlators}

On the other hand, in the case of a correlator involving $2n+1$ twisted Wilson loops we have
\begin{align}
\label{W2n1}
\langle W_{\alpha_1}W_{\alpha_2}\dots W_{\alpha_{2n+1}}^{\dagger} \rangle = \frac{1}{N^{2n+1}}\sum_{k_1=0}^{\infty}\dots\sum_{k_{2n+1}=0}^{\infty}\left(\prod_{i=1}^{2n+1}\frac{1}{k_i!}\left(\frac{\lambda}{2N}\right)^{\frac{k_i}{2}}\right)\,\langle A_{\alpha_1,k_1}\dots A_{\alpha_{2n+1},k_{2n+1}}^{\dagger} \rangle\,.
\end{align}
Since, as recalled in Section \ref{matrixmodel}, in the large-$N$ limit it holds that
\begin{align}
& \langle\, \mathcal{P}_{\alpha_1,k_1}\,\mathcal{P}_{\alpha_2,k_2} \dots \mathcal{P}_{\alpha_{2n},k_{2n}}\,\mathcal{P}_{\alpha_{2n+1},k_{2n+1}}^{\dagger} \rangle \simeq \nonumber \\[0.5em]
& \left(\frac{\delta_{\alpha_1+\alpha_2,\alpha_3}}{\sqrt{M}N}\,\textsf{d}_{k_1}^{(\alpha_1)}\,\textsf{d}_{k_2}^{(\alpha_2)}\,\textsf{d}_{k_3}^{(\alpha_3)}\right)\left(\delta_{\alpha_4,\alpha_5}\,\textsf{D}_{k_4,k_5}^{(\alpha_4)}\right) \dots   \left(\delta_{\alpha_{2n},\alpha_{2n+1}}\,\textsf{D}_{k_{2n},k_{2n+1}}^{(\alpha_{2n})}\right) +   \text{other Wick contractions}\,,
\label{2n+1ptP}
\end{align}
the correlator \eqref{W2n1} becomes in the planar limit
\begin{align}
& \langle\, W_{\alpha_1}\,W_{\alpha_2} \dots W_{\alpha_{2n}}W_{\alpha_{2n+1}}^{\dagger} \,\rangle \, \simeq \nonumber \\
&   V_{\alpha_1,\alpha_2,\alpha_3}\prod_{j=2}^{n}\left(\frac{\delta_{\alpha_{2j},\alpha_{2j+1}}}{N^2}\sum_{q=2}^{\infty}\sum_{p=2}^{\infty}I_q(\sqrt{\lambda})I_p(\sqrt{\lambda}) \, \sqrt{q\,p}\, \textsf{D}_{q,p}^{(\alpha_{2j})}\right)\, +   \, \text{other Wick contractions}\,,
\label{eq:W2n1LargeN}
\end{align}
with
\begin{align}
& V_{\alpha_1,\alpha_2,\alpha_3} \equiv  \frac{\delta_{\alpha_1+\alpha_2,\alpha_3}}{\sqrt{M}N^4}\left[ \, \prod_{p=1}^{3}\left( \sum_{k=1}^{\infty}\sum_{\ell=1}^{\infty}I_{2k}(\sqrt{\lambda})\sqrt{(2k)\,(2\ell)}\,\textsf{D}^{(\alpha_p)}_{2k,2\ell}\right)  + \right. \nonumber \\[0.5em]
& \left. \sum_{\sigma \in \mathcal{Q}_3}\left[\left(\sum_{k=1}^{\infty}\sum_{\ell=1}^{\infty}I_{2k}(\sqrt{\lambda})\sqrt{(2k)\,(2\ell)}\,\textsf{D}^{(\alpha_{\sigma(1)})}_{2k,2\ell}\right) \right. \right. \left(  \sum_{k=1}^{\infty}\sum_{\ell=1}^{\infty}I_{2k+1}(\sqrt{\lambda})\sqrt{(2k+1)\,(2\ell+1)}\,\textsf{D}^{(\alpha_{\sigma(2)})}_{2k+1,2\ell+1} \right) \nonumber \\[0.5em]
& \left. \left. \left(  \sum_{k=1}^{\infty}\sum_{\ell=1}^{\infty}I_{2k+1}(\sqrt{\lambda})\sqrt{(2k+1)\,(2\ell+1)}\,\textsf{D}^{(\alpha_{\sigma(3)})}_{2k+1,2\ell+1} \right) \right]\,  \right],
\label{eq:vertez123}
\end{align}
where the set of permutations $\mathcal{Q}_3$ is given by expression \eqref{setQ3}.
If we turn off the interaction action, the expression \eqref{eq:W2n1LargeN}
becomes
\begin{align}
\mathcal{N}^{odd}(\alpha_1,\alpha_2,\dots,\alpha_{2n+1})\,\frac{\lambda^{3/2}}{8\sqrt{M}N^4}I_1(\sqrt{\lambda})\Big[I_1(\sqrt{\lambda})^2+3I_2(\sqrt{\lambda})^2\Big]\left(\frac{\sqrt{\lambda}}{2\,N^2}\right)^{n-1}\left[I_1(\sqrt{\lambda})I_2(\sqrt{\lambda})\right]^{n-1} \, \ ,
\end{align}
where
\begin{align}
\mathcal{N}^{odd}(\alpha_1,\alpha_2,\dots,\alpha_{2n+1}) \equiv \delta_{\alpha_1+\alpha_2,\alpha_3}\,\delta_{\alpha_4,\alpha_5}\dots \delta_{\alpha_{2n},\alpha_{2n+1}} \, +    \, \text{other Wick contractions}\,,
\end{align}
and it counts the number of non-trivial Wick contractions. 
We consider the ratio
\begin{align}
\frac{\langle\, W_{\alpha_1}\,W_{\alpha_2}\dots W_{\alpha_{2n}}\,W_{\alpha_{2n+1}}^\dagger\rangle}{\mathcal{N}^{odd}(\alpha_1,\alpha_2,\dots\,,\alpha_{2n+1})\,\sqrt{M}\,W^{(3)}_{conn}(\lambda)\,(W^{(2)}_{conn}(\lambda))^{n-1}}\, \equiv \,
1+\Delta w^{(\alpha_1,\alpha_2,\dots, \alpha_{2n})}(M,\lambda)  \ ,
\end{align}
where $W_{conn}^{(2)}(\lambda)$ is defined in \eqref{Wconnected} while $W_{conn}^{(3)}(\lambda)$ in \eqref{eq:Z3connZM}. By construction the above ratio, when we turn off the interaction action, is equal to $1$. We evaluate it for large values of $\lambda$ and we find
\begin{align}
& 1+\Delta w^{(\alpha_1,\alpha_2,\dots, \alpha_{2n})}(M,\lambda) \,\underset{\lambda\rightarrow\infty}{\sim} \, -\frac{1}{\mathcal{N}^{odd}(\alpha_1,\alpha_2,\cdots,\alpha_{2n+1})} \nonumber \\
& \left[\frac{\delta_{\alpha_1+\alpha_2,\alpha_3}}{8}\prod_{i=1}^{3}\frac{\mathcal{I}_0(s_{\alpha_i})}{\sqrt{s_{\alpha_i}}}\prod_{j=2}^{n}\left(\delta_{\alpha_{2j},\alpha_{2j+1}}\frac{\mathcal{I}_0(s_{\alpha_{2j}})^2}{4\,s_{\alpha_{2j}}}\right) + \text{other Wick contractions} \right] + O\left(\frac{1}{\sqrt{\lambda}}\right)\,,
\label{oddFinal}
\end{align}
where we used the large-$\lambda$ behaviour of the 2- and 3- point correlators given in \eqref{eq:WWZMfinal} and \eqref{WWWabcSTRONG}, respectively.

\subsubsection{Mixed correlators}
In order to complete the analysis of the Wilson loop correlators in the planar limit of the $\mathbb{Z}_M$ quiver gauge theory, let us finally consider the following correlation function with $n$ untwisted Wilson loops and $m$ twisted Wilson loops 
\begin{align}
\label{eq:GeneralMixed}
& \langle\, \underbrace{W_0\, \cdots W_0}_{n}\, W_{\alpha_1}\, \cdots W_{\alpha_m}^{\dagger}  \,\rangle \, \ = \nonumber \\
& \frac{1}{N^{n+m}}\sum_{k_1=0}^{\infty}\dots\sum_{k_{n+m}=0}^{\infty}\left[\prod_{i=1}^{n+m}\frac{1}{k_i!}\left(\frac{\lambda}{2N}\right)^{\frac{k_i}{2}}\right]\langle A_{0,k_1}\dots A_{0,k_{n}}\hat{A}_{\alpha_1,k_{n+1}}\dots \hat{A}_{\alpha_{m},k_{n+m}}^{\dagger} \rangle \,.
\end{align}
For each of the untwisted Wilson loops we perform the change of basis in \eqref{Ahatoperators}, namely $A_{0,k_i}\,\mapsto\,\hat{A}_{0,k_i} + \sqrt{M}\,t_{k_i}$. Then, exploiting the properties of the $\mathcal{P}_{\alpha,k}$ operators at large-$N$, generalizing the procedure shown in Section \ref{mixed3pt}, it is easy to see that the leading contribution of \eqref{eq:GeneralMixed}  is due only to the term proportional to the product of the $t_{k_i}$. Therefore we find
\begin{align}
\langle\, \underbrace{W_0\, \cdots W_0}_{n}\, W_{\alpha_1}\, \cdots W_{\alpha_m}^{\dagger}  \,\rangle \,  \simeq \, \Big(\sqrt{M}\,\langle W \rangle_0\Big)^n \langle W_{\alpha_1}\, \cdots \, W_{\alpha_m}^{\dagger}\rangle \,,
\end{align}
where the correlator involving only twisted Wilson loops is given by the expression \eqref{evenFinal} or the expression \eqref{oddFinal} for $m$ even or odd respectively.

\section{Conclusions}
\label{sec: conclusions}

The main result of this paper is the systematic computation of Wilson loop $n$-point correlators in the planar limit of $\mathbb{Z}_M$ quiver gauge theories, which led us to find an exact expression for these correlation functions valid for every value of the 't Hooft coupling. In particular, we considered the reducible Wilson loop correlators, since the irreducible ones vanish in the planar limit. It would be interesting to extend our analysis also beyond the leading planar approximation, thus including the discussion of irreducible correlators, but it seems to be very intricate to find a precise structure for these correlation functions in this regime (to this regard the formulas in Appendix E of \cite{Billo:2022xas} could be helpful). Non-planar corrections to other observables in the $\mathbb{Z}_2$ quiver theory have been, for instance, recently investigated in \cite{Beccaria:2023kbl}.

Furthermore, we managed to derive the leading order of the strong coupling expansion of these $n$-point correlators in expressions \eqref{evenFinal} (for even $n$) and \eqref{oddFinal} (for odd $n$), which show that the non-trivial contribution of each twisted Wilson loop is captured by a remarkable simple rule in the planar limit for $\lambda\rightarrow\infty$, namely
\begin{align}
W_{\alpha_i} \rightarrow -\frac{\mathcal{I}_0(s_{\alpha_i})}{2\sqrt{s_{\alpha_i}}}\, .
\end{align}
A similar pattern was observed in \cite{Billo:2022fnb} for the 3-point functions among twisted chiral operators. In particular, for the $\mathbb{Z}_2$ case (i.e. only one twisted sector), we recover a large-$N$ factorization analogous to the structure exhibited by correlators among Wilson loops of $\mathcal{N}=4$ SYM in the 't Hooft limit, i.e. 
\begin{align}
\mathcal{N}=4 \, \ \ \text{SYM:}\ \ \ \ \ \ \ \ \ \ \ \ \ \ \ \ \ \langle W\dots W \rangle_0 \simeq (\langle W\rangle_0)^n\, \, ,
\end{align}
\begin{align}
\mathbb{Z}_2\ \ \text{theory:}\ \ \ \ \ \ \ \ \ \ \ \ \ \langle\, \underbrace{W_1\,W_1\cdots W_1}_{2n} \,\rangle \simeq (\langle W_1W_1 \rangle)^n \,,
\end{align}
which shows that some properties of the planar limit of $\mathcal{N}=4$ are inherited also by a proper combination of twisted observables of the $\mathbb{Z}_2$ quiver theory.

In all our analysis a crucial role has been played by the relation \eqref{G02pi} that provides an analytic expression for the generating function $G^{(0)}(s_{\alpha},-2\pi)$. From a purely mathematical point of view, this expression has been obtained by extending the analysis performed in \cite{Beccaria:2023kbl} to a $M$-nodes quiver gauge theory. Moreover, since the generating function $G^{(0)}(s_{\alpha},x)$ is not associated to any specific observable, the relation \eqref{G02pi} could turn out to be useful also in the evaluation of the strong coupling regime of other type of correlators with operators belonging to conjugated twisted sectors, e.g. $\langle W_{\alpha}\,T_{k,\alpha}^{\dagger}\rangle$  (where $T_{k,\alpha}$ denotes the chiral single trace twisted operators of \cite{Billo:2022fnb}). We leave the study of these quantities for future work.

It would also be interesting to examine the subleading corrections in the coupling in the planar limit of the theory, as already done for 2- and 3- point functions of chiral scalar operators, the v.e.v of a Wilson loop and the free energy \cite{Beccaria:2022ypy,Billo:2022lrv}. In order to determine these contributions, the knowledge of the subleading orders of the generating function $G(s_{\alpha},\ell,x)$, defined in \eqref{defG}, are required.

Finally, we stress the fact that these correlators, analysed in the large-$N$ 't Hooft limit in the strong coupling regime, should be described and, possibly, computed from the dual supergravity theory through the AdS/CFT correspondence. Even though, to the best of our knowledge, a holographic prediction for these correlation functions is presently unavailable, the expressions that we found could be regarded as a QFT-side prediction and, therefore, constitute a natural starting point for a future holographic investigation. We plan to carry out this analysis in the future.

\vskip 1cm
\noindent {\large {\bf Acknowledgments}}
\vskip 0.2cm
We are very grateful to A. Lerda, M. Frau and M. Billò for many important discussions and for reading and commenting on the draft of our article. We are also grateful to G. P. Korchemsky  for very useful discussions. This research is partially supported by the MUR PRIN contract 2020KR4KN2 ``String Theory as a bridge between Gauge Theories and Quantum Gravity'' and by the INFN project ST\&FI ``String Theory \& Fundamental Interactions''. 
\vskip 1cm

\appendix

\section{Identities between Bessel functions}
\label{appendix:A}

In this Appendix we prove some identities between the modified Bessel functions of the first kind that we have used in the derivation of the exact expression of $n$-point Wilson loop correlators in the planar limit. We start from the following identities
\begin{align}
 & \sum_{k=1}^{\infty}(2k+1)I_{2k+1}(z) = \frac{z}{2}I_2(z)  \, , \label{Iodd}\\
 &  \sum_{k=1}^{\infty}(2k)I_{2k}(z) = \frac{z}{2}I_1(z)\, . \label{Ieven}
\end{align}
We begin from the first one. We rewrite the modified Bessel function $I_{2k+1}$ in terms of a Bessel function of the first kind $J_{2k+1}$.
Then, substituting the resulting expression in \eqref{Iodd}, we get
\begin{align}
-i\sum_{k=1}^{\infty}(-1)^k(2k+1)J_{2k+1}(iz)\, .
\end{align}
At this point we can exploit the identity (A.20a) of \cite{Billo:2022fnb}, obtaining
\begin{align}
-i\sum_{k=1}^{\infty}(-1)^k(2k+1)J_{2k+1}(iz) = -\frac{z}{2}J_{2}(iz) = \frac{z}{2} I_{2}(z)\, .
\end{align}
This concludes the proof of \eqref{Iodd}. The proof of \eqref{Ieven} is analogous with the only difference that we have to use  identity (A.20b) of \cite{Billo:2022fnb}.

We now formally derive another useful identity
\begin{align}
\label{idIk}
\sum_{k=2}^{\infty} \, I_{k}(\sqrt{\lambda})^2\, k = \frac{\sqrt{\lambda}}{2}I_{1}(\sqrt{\lambda})I_{2}(\sqrt{\lambda})\, .
\end{align}
In order to prove it, we firstly consider 
\begin{align}
\sum_{k=2}^{\infty} \, I_{k}(\sqrt{x})I_{k}(\sqrt{y})\, k
\end{align}
and then we will take the particular limit $\sqrt{x}\rightarrow\sqrt{y}\rightarrow\sqrt{\lambda}$.
As a first step we split the sum over even and odd contributions
\begin{align}
\sum_{k=2}^{\infty} \, I_{k}(\sqrt{x})I_{k}(\sqrt{y})\, k = \sum_{k=1}^{\infty} (2k)\,I_{2k}(\sqrt{x})I_{2k}(\sqrt{y})\,+\, \sum_{k=1}^{\infty} (2k+1)\,I_{2k+1}\, (\sqrt{x})I_{2k+1}(\sqrt{y})\, .
\end{align}
Now we exploit the following identities
\begin{align}
& \sum_{k=1}^{\infty} 2k\,I_{2k}(\sqrt{x})I_{2k}(\sqrt{y}) = \frac{\sqrt{x\,y}}{2(x-y)}\left(\sqrt{x}I_2(\sqrt{x})I_1(\sqrt{y})-\sqrt{y}I_{2}(\sqrt{y})I_1(\sqrt{x})\right)\, , \\
& \sum_{k=1}^{\infty} (2k+1)\,I_{2k+1}(\sqrt{x})I_{2k+1}(\sqrt{y}) = \frac{\sqrt{x\,y}}{2(x-y)}\left(\sqrt{x}I_1(\sqrt{x})I_0(\sqrt{y})-\sqrt{y}I_{1}(\sqrt{y})I_0(\sqrt{x})\right) -I_1(\sqrt{x})I_1(\sqrt{y})\, ,
\end{align}
which can be proven by considering the analogous identities valid for the Bessel functions of the first kind
\begin{align}
& \sum_{k=1}^{\infty} 2k\,J_{2k}(i\sqrt{x})J_{2k}(i\sqrt{y})  = \frac{i\,\sqrt{x\,y}}{2(x-y)}\left(\sqrt{x}J_2(i\sqrt{x})J_1(i\sqrt{y})-\sqrt{y}J_{2}(i\sqrt{y})J_1(i\sqrt{x})\right)\, , \\
& \sum_{k=1}^{\infty} (2k+1)\,J_{2k+1}(i\sqrt{x})J_{2k+1}(i\sqrt{y})  = \frac{i\,\sqrt{x\,y}}{2(x-y)}\left(\sqrt{x}J_1(i\sqrt{x})J_0(i\sqrt{y})-\sqrt{y}J_{1}(i\sqrt{y})J_0(i\sqrt{x})\right)\nonumber \\
& \ \ \ \ \  \ \  \ \  \ \  \ \  \ \ \  \ \ \ \  \  \ \ \  \ \ \  \   \ \ \  \ \  \ \ \ \ \ \  \  \ \ \ \  \ \ \ \   -J_1(i\sqrt{x})J_1(i\sqrt{y}) \, \ .
\end{align}
Thus we are left with
\begin{align}
\sum_{k=2}^{\infty} \, I_{k}(\sqrt{x})I_{k}(\sqrt{y})\, k & =  \frac{\sqrt{x\,y}}{2(x-y)}\left(\sqrt{x}I_2(\sqrt{x})I_1(\sqrt{y})-\sqrt{y}I_{2}(\sqrt{y})I_1(\sqrt{x})\right) \notag \\ & + \frac{\sqrt{x\,y}}{2(x-y)}\left(\sqrt{x}I_1(\sqrt{x})I_0(\sqrt{y})-\sqrt{y}I_{1}(\sqrt{y})I_0(\sqrt{x})\right) -I_1(\sqrt{x})I_1(\sqrt{y})\, .
\end{align}
Now we use the following identity between the Bessel functions
\begin{align}
\label{idI2}
I_{0}(\sqrt{x})=I_2(\sqrt{x})+\frac{2}{\sqrt{x}}I_1(\sqrt{x})\,,
\end{align}
so that the previous expression becomes
\begin{align}
\sum_{k=2}^{\infty} \, I_{k}(\sqrt{x})I_{k}(\sqrt{y})\, k & =  \frac{\sqrt{x\,y}}{2(\sqrt{x}+\sqrt{y})}\left(I_2(\sqrt{x})I_1(\sqrt{y})+I_{2}(\sqrt{y})I_1(\sqrt{x})\right)\, .
\end{align}
Thus we equal the arguments of the Bessel functions and we set them to $\sqrt{\lambda}$ obtaining
\begin{align}
\sum_{k=2}^{\infty} \, I_{k}(\sqrt{\lambda})^2\, k & =  \frac{\sqrt{\lambda}}{2}I_1(\sqrt{\lambda})I_2(\sqrt{\lambda})\, .
\end{align}

\section{The coefficients \texorpdfstring{$w_{n,m}^{(\ell)}(s_{\alpha})$}{} and their generating functions}
\label{app:Galpha}
In this appendix we derive the strong coupling expansion of the coefficients $w_{n,m}^{(\ell)}(s_{\alpha})$ and some properties of the corresponding generating functions. Finally, using both analytical and numerical techniques, we provide an argument for the validity of the relation \eqref{G02pi}.

\subsection{The  \texorpdfstring{$\textsf{X}$}{}-matrix as an operator}
We rewrite the matrix elements \eqref{Xmatrix} as
\begin{align}
\label{2Xmatrix}
\textsf{X}_{n,m} = 2\,(-1)^{\frac{n+m +2\,nm}{2}}\sqrt{n\,m} \int_0^{\infty} \frac{dt}{t}\,\chi(t)\,J_{n}\left(\frac{\sqrt{\lambda}\,t}{2\pi}\right)J_{m}\left(\frac{\sqrt{\lambda}\,t}{2\pi}\right) \, ,
\end{align}
where we introduced the symbol function
\begin{align}
\label{eq:ChiSymbol}
\chi(t) = -\frac{4\,\text{e}^t}{(\text{e}^t-1)^2} = -\sinh\left(\frac{t}{2}\right)^{-2} \, .
\end{align}
Then, with the aim to simplify the notation, we rescale the 't Hooft coupling introducing
\begin{align}
\label{g}
g =  \frac{\sqrt{\lambda}}{4\pi} 
\end{align}
and we perform the change of variable
\begin{align}
\sqrt{x} = 2gt \,  ,
\end{align}
this way the matrix elements \eqref{2Xmatrix} become
\begin{align}
\label{3Xmatrix}
\textsf{X}_{n,m} = (-1)^{\frac{n+m+2nm}{2}} \sqrt{n\,m}\int_{0}^{\infty} \frac{dx}{x}\, \chi\left(\frac{\sqrt{x}}{2g}\right) J_{n}(\sqrt{x})\,J_{m}(\sqrt{x}) \, .
\end{align}
We then introduce the set of functions
\begin{align}
\label{psik}
\psi_k(x)=(-1)^{\frac{k}{2}(k-1)}\sqrt{k}\,\frac{J_k(\sqrt{x})}{\sqrt{x}}  \,,
\end{align}
with $k=1,2,\dots $ Starting from them we can construct two orthonormal basis, namely
\begin{subequations}
\begin{equation}
 \langle \psi_{2k} | \psi_{2\ell} \rangle = \int_{0}^{\infty}\,dx\,\psi_{2k}(x)\psi_{2\ell}(x)=\delta_{k,\ell} \,   ,
\end{equation}
\begin{equation}
 \  \langle \psi_{2k+1} | \psi_{2\ell+1} \rangle = \int_{0}^{\infty}\,dx\,\psi_{2k+1}(x)\psi_{2\ell+1}(x)=\delta_{k,\ell} \, ,
\end{equation}
\label{OddBasis}
\end{subequations}
This way we can regard the non trivial elements\footnote{We remember that by definition (see equation \eqref{XEvenOdd}) the entries of the $\textsf{X}$-matrix with opposite parity vanish.} of the $\textsf{X}$-matrix \eqref{3Xmatrix} as an integral representation of the operator $\textsf{X}$ acting on the basis \eqref{OddBasis}, namely 
\begin{align}
\label{Xoperator}
\textsf{X}_{k,\ell} = \langle \psi_{k} | \textsf{X} |\psi_\ell \rangle   \, \ .
\end{align} 
In the following part of this Appendix it will be very useful to think about the $\textsf{X}$-matrix in a more abstract way, namely as an operator acting among infinite dimensional vector spaces. This in turn will allow to understand some of its properties at strong coupling. We warn the reader that we will employ the same symbol, i.e.  $\textsf{X}$, to denote both the operator and its infinite dimensional representation. 
\subsection{The coefficients  \texorpdfstring{$w_{n,m}^{(\ell)}(s_{\alpha})$}{}}
In this section we aim to study the main properties of the coefficients  $w_{n,m}^{(\ell)}(s_{\alpha})$ introduced in \eqref{wnmalpha} and whose explicit expression reads 
\begin{align}
\label{startwnm}
w_{n,m}^{(\ell)}(s_{\alpha}) =
\langle (x\partial_x)^n\phi^{(\ell)}(x) \Big|\frac{s_{\alpha}\textsf{X}}{1-s_{\alpha}\textsf{X}}\Big|(x\partial_x)^m\phi^{(\ell)}(x) \rangle   \,  ,
\end{align}
where $\phi^{(\ell)}(x) = J_{\ell}(\sqrt{x})$.
We observe that \eqref{startwnm} can be obtained starting from the expression (C.1) of \cite{Beccaria:2023kbl} simply by a rescaling of the symbol function \eqref{eq:ChiSymbol}, namely
\begin{align}
\label{chiRescaling}
    \chi(x) \mapsto s_{\alpha}\chi(x) \, .
\end{align}
Therefore \eqref{startwnm} constitutes the generalization, valid to each arbitrary value of $s_{\alpha}$, of the coefficients $w_{nm}$ introduced in \cite{Beccaria:2023kbl}, i.e. $ w_{n,m}^{(\ell)}(1) \equiv w_{nm}$.

Let us now move to analyse the properties of the coefficients \eqref{startwnm} valid for any value of the coupling $g$.
Following the same steps performed in Appendix C of \cite{Beccaria:2023kbl}, one can firstly show that the coefficients $w_{n,m}^{(\ell)}(s_{\alpha})$ satisfy the differential equation
\begin{align}
\left(\frac{1}{2}g\partial_{g}-1\right)w_{n,m}^{(\ell)}(s_{\alpha}) = \frac{1}{4}w_{0,n}^{(\ell)}(s_{\alpha})w_{0,m}^{(\ell)}(s_{\alpha})+ w_{n+1,m}^{(\ell)}(s_{\alpha})+w_{n,m+1}^{(\ell)}(s_{\alpha}) \,,
\label{eqwnm}
\end{align}
which implies that only a subset of the initial coefficients \eqref{startwnm} is independent, namely the ones given by the $w_{0,n}^{(\ell)}(s_{\alpha})$ with $n=0,2,4,\dots\,$ . As a matter of fact all the others can be obtained by exploiting equation \eqref{eqwnm}, for example
\begin{align}
& w_{0,1}^{(\ell)}(s_{\alpha}) = -\frac{1}{2}\left(1+\frac{1}{4}w_{0,0}^{(\ell)}(s_{\alpha})-\frac{1}{2}g\partial_g\right)w_{0,0}^{(\ell)}(s_{\alpha})\, , \\[0.5em]
& w_{1,1}^{(\ell)}(s_{\alpha})= -\left(1+\frac{1}{4}w^{(\ell)}_{0,0}(s_{\alpha})-\frac{1}{2}g\partial_{g}\right)w^{(\ell)}_{0,1}(s_{\alpha})-w^{(\ell)}_{0,2}(s_{\alpha}) \,  .
\end{align}

Then one can also show that the coefficients \eqref{startwnm} satisfy the equations
\begin{subequations}
\begin{equation}
\partial_{g}\, w_{0,n}^{(\ell)}(s_{\alpha}) = -8g \int_{0}^{\infty} dz\, z^2 q_0^{(\ell)}\,(z,g,s_{\alpha})\,q_n^{(\ell)}(z,g,s_{\alpha})\,\partial_{z}(s_{\alpha}\chi(z))\,, \label{inteq}
\end{equation}
\begin{equation}
q_{n+1}^{(\ell)}(z,g,s_{\alpha}) = -\frac{1}{4}q_0^{(\ell)}(z,g,s_{\alpha})w_{0,n}^{(\ell)}(s_{\alpha})+\frac{1}{2}g\partial_{g}q_n^{(\ell)}(z,g,s_{\alpha}) \label{receq} \,,    
\end{equation}
\label{eq:diff}
\end{subequations}
where the function $q_0^{(\ell)}(z,g,s_{\alpha})$ is a solution of the differential equation
\begin{align}
\label{eqQ0}
 & [(g\partial_g)^2+4(gz)^2-\ell^2+(1-g\partial_g)w_{0,0}^{(\ell)}(s_{\alpha})]q_0^{(\ell)}(z,g,s_{\alpha}) = 0\,,
\end{align}
subject to the boundary condition at weak coupling $q_0^{(\ell)}(z,g,s_{\alpha}) = J_{\ell}(2gz) + O(g^{2\ell+1})$. 
We observe that applying several times \eqref{receq} we can express $q_n^{(\ell)}(z,g,s_{\alpha})$ in terms of $q_0^{(\ell)}(z,g,s_{\alpha})$ and the coefficients $w_{0,n}^{(\ell)}(s_{\alpha})$, namely 
\begin{align}
\label{qn}
q_n^{(\ell)}(z,g,s_{\alpha}) = -\frac{1}{4}\sum_{j=0}^{n-1}\left(\frac{1}{2}\right)^{j}(g\partial_{g})^j(q_0^{(\ell)}(z,g,s_{\alpha})w_{0,n-1-j}^{(\ell)}(s_{\alpha})) + \left(\frac{1}{2}\right)^{n}(g\partial_{g})^nq_0^{(\ell)}(z,g,s_{\alpha}) \,  .
\end{align}

Henceforth we focus on the strong coupling regime and we show how the iterative use of the equations \eqref{inteq} and \eqref{qn} permits to determine the strong coupling expansion of the coefficients $w^{(\ell)}_{0,n}(s_{\alpha})$ with $n=0,2,4,\dots$\, . Following \cite{Beccaria:2023kbl} we assume that for large-$g$ they scale as
\begin{align}
\label{Strongwnm}
w_{0,n}^{(\ell)}(s_{\alpha}) = \sum_{i \,\geq \, 0}w_{0,n}^{(\ell),(i)}(s_{\alpha})\,g^{n+1-i}\,,
\end{align}
where the index $i$ labels the order of the expansion.
Although this analysis could be performed in full generality, for the scope of this article it is enough to determine the leading term of \eqref{Strongwnm}, i.e. the coefficient $w_{0,n}^{(\ell),(0)} \equiv \omega^{(\ell)}_{0,n}(s_{\alpha})$. Therefore, in the following, we will mainly focus on this quantity and we will just briefly comment on the computation of the subleading terms of \eqref{Strongwnm}.

Let us start considering the $n=0$ case. The coefficient $\omega_{0,0}^{(\ell)}(s_{\alpha})$ can be computed applying the same procedure discussed in \cite{Belitsky:2020qir} with the only difference that the symbol $\chi(z)$ must be rescaled as in \eqref{chiRescaling}, this way we find
\begin{align}
\label{omega00}
\omega_{0,0}^{(\ell)}(s_{\alpha}) = -4\int_0^{\infty} \frac{dz}{\pi}\log(1-s_{\alpha}\chi(z)) = 4\,\mathcal{I}_0(s_{\alpha}) \, \ , 
\end{align}
where we introduced the function
\begin{align}
\label{Inalpha}
\mathcal{I}_n(s_{\alpha}) = \int_{0}^{\infty}\frac{dz}{\pi}z^{1-2n}\partial_{z}\log(1-s_{\alpha}\chi(z)) \,.
\end{align}
Then let us consider equation \eqref{eqQ0}. A solution to this equation can be found applying the semi-classical methods of \cite{Belitsky:2020qrm,Belitsky:2020qir} and, at the leading order of the expansion, can be taken of the form $\mathcal{C}(z,s_{\alpha})J_{\ell}(2gz)$, where $\mathcal{C}(z,s_{\alpha})$ is an arbitrary function of $z$ and $s_{\alpha}$. Then, $\mathcal{C}(z,s_{\alpha})$ is fixed by evaluating for large-$g$ the integral on the r.h.s. of \eqref{inteq}  and demanding that is equal to \eqref{omega00}. This way, after some mathematical steps, we conclude that
\begin{align}
\mathcal{C}(z,s_{\alpha}) = \frac{1}{\sqrt{1-s_{\alpha}\chi(z)}}\, \ .
\end{align}
Let us now  construct the solution $q_0(z,g,s_{\alpha})$ valid for any order of the large-$g$ expansion. Firstly we observe that,
for large $gz$, the Bessel function $J_{\ell}(2gz)$ admits the following expansion
\begin{align}
J_{\ell}(2gz) \underset{g z\rightarrow\infty}{\sim} \frac{1}{\sqrt{2\pi\,g\,z}}\left((-1)^{\frac{\ell}{2}(\ell-1)}\sin(2gz)+(-1)^{\frac{\ell}{2}(\ell-3)}\cos(2gz)\right)\, ,
\end{align}
which in turn suggests to consider the following ansatz for the general solution
\begin{align}
& q_0^{(\ell)}(z,g,s_{\alpha}) = \nonumber \\
& \frac{1}{\sqrt{1-s_{\alpha}\chi(z)}}\frac{1}{\sqrt{2\pi\,g\,z}}\left[\biggl(a_{0,0}^{(\ell)}+\sum_{k=1}^{\infty}\frac{a_{0,k}^{(\ell)}(z)}{(gz)^k}\biggr)\text{sin}(2gz)+\biggl(b_{0,0}^{(\ell)}+\sum_{k=1}^{\infty}\frac{b_{0,k}^{(\ell)}(z)}{(gz)^k}\biggr)\text{cos}(2gz)\right]\ \, ,
\label{f0ansatz}
\end{align}
where
\begin{align}
 a_{0,0}^{(\ell)} = (-1)^{\frac{\ell}{2}(\ell-1)}\,,\ \ \ \ b_{0,0}^{(\ell)} = (-1)^{\frac{\ell}{2}(\ell-3)}\,.
\end{align}
Thus, by construction, the leading term of \eqref{f0ansatz} coincides with the leading term of $\mathcal{C}(z,s_{\alpha})J_{\ell}(2gz)$, while the subleading contributions, i.e. $a^{(\ell)}_{0,k}(z)$ and $b^{(\ell)}_{0,k}(z)$ with $k\geq 1$, are determined by requiring that \eqref{f0ansatz} satisfies \eqref{eqQ0}. This way we construct a solution valid at any order of the strong coupling expansion. Then, the subleading coefficients of the expansion \eqref{Strongwnm} can be determined substituting the expression \eqref{f0ansatz} in \eqref{inteq} and imposing the equality at each order of the large-$g$ expansion. This leads to
\begin{align}
w_{0,0}^{(\ell)}(s_{\alpha}) = & \,4g\,\mathcal{I}_0(s_{\alpha}) + (2\ell-1) -\frac{(2\ell-1)(2\ell-3)\mathcal{I}_1(s_{\alpha})}{8g} - \frac{(2\ell-1)(2\ell-3)\mathcal{I}_1(s_{\alpha})^2}{16g^2} \nonumber \\
& -\frac{(2\ell-1)(2\ell-3)(16\mathcal{I}_1(s_{\alpha})^3-5\mathcal{I}_2(s_{\alpha})-8\mathcal{I}_2(s_{\alpha})\ell+4\mathcal{I}_2(s_{\alpha})\ell^2)}{512g^3} + O\left(\frac{1}{g^4}\right)\,.
\end{align}

Now we consider the cases with $n \geq 1$ and we just focus on the leading order of the large-$g$ expansion. The expression \eqref{f0ansatz} and the recursion relation \eqref{qn} suggest the following ansatz for the functions $q_n(z,g,s_{\alpha})$ with $n \geq 1$
\begin{align}
\label{qnansatz}
q_{n}(z,g,s_{\alpha}) = \frac{1}{\sqrt{1-s_{\alpha}\chi(z)}}\frac{1}{\sqrt{2\pi\,g\,z}}\left[a_n(z,g)\sin(2gz) + b_n(z,g)\cos(2gz)\right] \, .
\end{align}
Then the iterative use of the relation \eqref{qn} and of equation \eqref{inteq} permits to determine the expressions of the remaining coefficients $\omega_{0,n}^{(\ell)}(s_{\alpha})$ with $n \geq 1$. For example for $n=1$ the relation \eqref{qn} reads
\begin{align}
q_1^{(\ell)}(z,g,s_{\alpha}) = \frac{1}{2}g \partial_{g}q_0^{(\ell)}(z,g,s_{\alpha})
-\frac{1}{4}q_0^{(\ell)}\omega^{(\ell)}_{0,0}\,g\,  .
\end{align}
Then, using \eqref{f0ansatz} we determine the coefficients $a_1(z,g)$ and $b_1(z,g)$ appearing in the large-$g$ expansion of $q_1(z,g,s_{\alpha})$, finding
\begin{align}
a_1(z,g) = -(a_{0,0}^{(\ell)}\,\mathcal{I}_0(s_{\alpha}) - b_{0,0}^{(\ell)}\,z)\,g\, , \, \ \ \ b_1(z,g) = -(b_{0,0}^{(\ell)}\,\mathcal{I}_0(s_{\alpha}) - a_{0,0}^{(\ell)}\,z)\,g \, , 
\end{align}
this way we completely fix the leading order expression of $q_1(z,g,s_{\alpha})$. Then we require that equation \eqref{inteq} is satisfied for the case at hand and we find
\begin{align}
\omega^{(\ell)}_{0,1}(s_{\alpha}) = -2\,\mathcal{I}_0(s_{\alpha})^2\,.
\end{align}
Let us also consider the $n=2$ case, this time \eqref{qn} reads
\begin{align}
q_2^{(\ell)}(z,g,s_{\alpha}) = -\frac{1}{4}\,q_0^{(\ell)}(z,g,s_{\alpha})\omega_{0,1}^{(\ell)}(s_{\alpha})\,g^2 - \frac{1}{8}\, g\partial_{g}(q_0^{(\ell)}(z,g,s_{\alpha})\omega_{0,0}^{(\ell)}\,g) - g^2z^2q_0^{(\ell)}(z,g,s_{\alpha}) \, , 
\end{align}
where, using the leading term of equation \eqref{eqQ0}, we wrote 
\begin{align}
(g\partial_g)^2q_0^{(\ell)}(z,g,s_{\alpha}) = -4(gz)^2q_0^{(\ell)}(z,g,s_{\alpha}) \, .
\end{align}
Similarly to the previous case, we determine the coefficients $a_2(z,g)$ and $b_2(z,g)$ and we demand that equation \eqref{inteq} is satisfied. We finally find
\begin{align}
\omega_{0,2}^{(\ell)}(s_{\alpha}) = \frac{1}{3}\left(2\,\mathcal{I}_0(s_{\alpha})^2-4\,\mathcal{I}_{-1}(s_{\alpha})\right) = -\frac{1}{3}\left(\mathcal{I}_0(s_{\alpha})\omega_{0,1}^{(\ell)}(s_{\alpha})+4\,\mathcal{I}_{-1}(s_{\alpha})\right) \, . 
\end{align}
Applying $n$ number of times this procedure we determine the general expressions
\begin{subequations}
\begin{align}
& \omega^{(\ell)}_{0,2n+1}(s_{\alpha}) = -\frac{1}{2(n+1)}\sum_{j=0}^{n}(-1)^j\, \mathcal{I}_{-j}(s_{\alpha})\, \omega^{(\ell)}_{0,2n-2j}(s_{\alpha})\,,  \label{wodd} \\
& \omega^{(\ell)}_{0,2n}(s_{\alpha}) = \frac{1}{2n+1} \left(\sum_{j=1}^{n} (-1)^{n-j-1} \, \mathcal{I}_{-n+j}(s_{\alpha})\, \omega^{(\ell)}_{0,2j-1}(s_{\alpha}) + 4(-1)^{n}\mathcal{I}_{-n}(s_{\alpha}) \right)\, . \label{weven}
\end{align}
\label{omegaLeading}
\end{subequations}
The relations \eqref{omegaLeading} permit to recursively determine the leading term of the  $w_{0,n}^{(\ell)}(s_{\alpha})$ coefficients, once are known the expressions of the integrals $\mathcal{I}_{-n}(s_{\alpha})$.

\subsection{The generating functions}
It is useful to introduce the generating functions for the coefficients $w_{0,n}^{(\ell)}(s_{\alpha})$ and $w_{n,m}^{(\ell)}(s_{\alpha})$, namely
\begin{align}
& G(s_{\alpha},\ell,x) = \sum_{n=0}^{\infty}\frac{w^{(\ell)}_{0,n}(s_{\alpha})}{(g\,x)^n} = g\, G^{(0)}(s_{\alpha},x) + G^{(1)}(s_{\alpha},\ell,x)+g^{-1}\,G^{(2)}(s_{\alpha},\ell,x) + \dots \,, \label{defG} \\
& G(s_{\alpha},\ell,x,y) = \sum_{n=0}^{\infty}\sum_{m=0}^{\infty}\frac{w^{(\ell)}_{n,m}(s_{\alpha})}{(gx)^n\,(gy)^m} = g\,G^{(0)}(s_{\alpha},x,y) + G^{(1)}(s_{\alpha},\ell,x,y) + g^{-1}G^{(2)}(s_{\alpha},\ell,x,y)+ \dots \,, \label{defGnm} 
\end{align}
where $G^{(0)}(s_{\alpha},x)$ denotes the generating function for the $\omega_{0,n}^{(\ell)}(s_{\alpha})$ coefficients, $G^{(0)}(s_{\alpha},x,y)$ the generating function for the $\omega^{(\ell)}_{n,m}(s_{\alpha})$ ones, while the other terms in the sums on the r.h.s. of the expressions \eqref{defG}-\eqref{defGnm} are the generating functions for the subleading contributions in the expansions of $w_{0,n}^{(\ell)}(s_{\alpha})$ and $w_{n,m}^{(\ell)}(s_{\alpha})$.

It is important to note that $G(s_{\alpha},\ell,x)$ and $G(s_{\alpha},\ell,x,y)$ are not independent. Indeed, generalizing the procedure discussed in \cite{Beccaria:2023kbl} for the case at hand, if we multiply both sides of equation \eqref{eqwnm}  by $(gx)^{-n}(gy)^{-m}$  and we sum over $n$ and $m$, we obtain
\begin{align}
\label{GGequation}
 (gx+gy+1)G(s_{\alpha},\ell,x,y) -\frac{1}{2}g\sum_{n=0}^{\infty}\sum_{m=0}^{\infty}(gx)^{-n}(gy)^{-m}\partial_g w^{(\ell)}_{n,m}(s_{\alpha}) = & -\frac{1}{4}G(s_{\alpha},\ell,x)G(s_{\alpha},\ell,y) \notag \\
 & +gx\,G(s_{\alpha},\ell,y)+gy\,G(s_{\alpha},\ell,x)\, .
\end{align}
Then we rewrite the second term on the l.h.s. as
\begin{align}
& -\frac{1}{2}g\sum_{n=0}^{\infty}\sum_{m=0}^{\infty}(gx)^{-n}(gy)^{-m}\partial_g w^{(\ell)}_{n,m}(s_{\alpha})  = -\frac{1}{2}g\partial_gG(s_{\alpha},\ell,x,y) \nonumber \\
& - \frac{1}{2}\sum_{n=0}^{\infty}\sum_{m=0}^{\infty}(n+m)(gx)^{-n}(gy)^{-m} w^{(\ell)}_{n,m}(s_{\alpha}) = -\frac{1}{2}g\partial_g G(s_{\alpha},\ell,x,y) +\frac{1}{2} (x\partial_x+y\partial_y)G(s_{\alpha},\ell,x,y)\, \ ,
\end{align}
substituting this expression back in \eqref{GGequation} we obtain an equation that connects $G(s_{\alpha},\ell,x,y)$ with $G(s_{\alpha},\ell,x)$, namely
\begin{align}
\label{GGequationFinal}
 (gx+gy+1)G(s_{\alpha},\ell,x,y) +\frac{1}{2}(x\partial_x+y\partial_y-g\partial_g)G(s_{\alpha},\ell,x,y) = & -\frac{1}{4}G(s_{\alpha},\ell,x)G(s_{\alpha},\ell,y) \notag \\
 & +gx\,G(s_{\alpha},\ell,y)+gy\,G(s_{\alpha},\ell,x)\, .
\end{align}
This way, once $G(s_{\alpha},\ell,x)$ is known, we can immediately determine $G(s_{\alpha},\ell,x,y)$. 
In particular, at the leading order of the large-$g$ expansion, the above relation \eqref{GGequationFinal} reads
\begin{align}
\label{GxyGxL}
    (x+y)G^{(0)}(s_{\alpha},x,y) = -\frac{1}{4}G^{(0)}(s_{\alpha},x)G^{(0)}(s_{\alpha},y)+x\,G^{(0)}(s_{\alpha},y)+y\,G^{(0)}(s_{\alpha},x)\, \ . 
\end{align}
We finally observe that setting $s_{\alpha}=1$ we recover equation (5.8) of \cite{Beccaria:2023kbl}.
\subsection{The integrals \texorpdfstring{$\mathcal{I}_{n}(s_{\alpha})$}{} and the generating function \texorpdfstring{$G^{(0)}(s_{\alpha},-2\pi)$}{}}
\label{subsec:Integrals}
The coefficients $\omega^{(\ell)}_{0,n}(s_{\alpha})$, and consequently the $\omega^{(\ell)}_{n,m}(s_{\alpha})$, can be written in terms of the function $\mathcal{I}_n({s_{\alpha}})$ defined in \eqref{Inalpha}. According to equation (4.46) of \cite{Beccaria:2022ypy} the expression of these integrals for $n \geq 1$ is
\begin{align}
\mathcal{I}_n(s_{\alpha}) = \frac{1}{\Gamma(2n-1)}\frac{(-1)^n}{(2\pi)^{2n-1}}\left[\psi^{2(n-1)}\left(\frac{\alpha}{M}\right)+\psi^{2(n-1)}\left(1-\frac{\alpha}{M}\right)-2\psi^{2(n-1)}(1)\right] \,,
\end{align}
where $M$ is the number of nodes of the quiver  and $\psi^{(m)}$ denotes the $m$-th derivative of the digamma function. However, here we need the general expression of the above integrals also for $n \leq 0$. For $s_{\alpha}=1$ it was found that \cite{Beccaria:2023kbl}
\begin{align}
\mathcal{I}_{n}\left(1\right) = 2\,(-1)^{n-1} (1-2^{2-2n}) \pi^{1-2n} \zeta (2 n-1)\, \ . 
\label{salpha1}
\end{align}
The case $n=1$ require some attention and can be obtained taking the limit of \eqref{salpha1}, namely
\begin{align}
\lim_{n \to 1} \mathcal{I}_{n}(1) = \frac{2\log(2)}{\pi} \,  . 
\end{align}
Although we are not aware of the general expression of the integrals $\mathcal{I}_n(s_{\alpha})$ valid for generic value of $s_{\alpha}$, in the specific cases in which  $s_{\alpha}=\frac{3}{4},\frac{1}{2},\frac{1}{4}$, 
relying on numerical results, we find\footnote{These cases are involved in the $\mathbb{Z}_{3,4,6}$ quivers respectively.}
\begin{subequations}
\begin{align}
& \mathcal{I}_{n}\left(\frac{1}{2}\right) = (-1)^{n-1} 2^{-2 n-1} \left(4^n-4\right)
   \left(4^n+2\right)\pi^{1-2n} \zeta (2 n-1)\,, \label{12} \\[1em]
& \mathcal{I}_{n}\left(\frac{3}{4}\right) = (-1)^n 3^{2 n-1}
   \left(9^{1-n}-1\right) (2 \pi
   )^{1-2 n} \zeta (2 n-1)\,, \label{34} \, \  \\[1em]
& \mathcal{I}_{n}\left(\frac{1}{4}\right) = (-1)^n\frac{1}{6}\left(6+3\cdot4^n+2\cdot9^n-36^n\right)(2\pi)^{1-2n}\zeta(2n-1)\,. \label{14}
\end{align}
\label{intset}
\end{subequations}
As for the $s_{\alpha}=1$ case the evaluation of the above expressions \eqref{intset} for $n=1$ is a bit subtle. However we can properly define these integrals taking the following limits 
\begin{align}
& \mathcal{I}_1\left(\frac{3}{4}\right) \equiv \lim_{n \to 1}\mathcal{I}_{n}\left(\frac{3}{4}\right) = \frac{3 \log (3)}{2 \pi }\, , \ \ \mathcal{I}_1\left(\frac{1}{2}\right) \equiv \lim_{n \to 1}\mathcal{I}_{n}\left(\frac{1}{2}\right) = \frac{3 \log (2)}{\pi }\, , \nonumber \\[0.5em]
&  \mathcal{I}_{1}\left(\frac{1}{4}\right) \equiv \lim_{n \to 1}\mathcal{I}_{n}\left(\frac{1}{4}\right) = \frac{1}{\pi}\left(2\log(2)+\frac{3}{2}\log(3)\right)\,.  
\end{align}
Crucially for us we find that for $n \leq 0$ we can express the integrals \eqref{intset} as a function of $\mathcal{I}_0(s_{\alpha})$, namely 
\begin{subequations}
\begin{align}
& \mathcal{I}_{n}\left(\frac{3}{4}\right) = \frac{3}{4}\mathcal{I}_0\left(\frac{3}{4}\right)\left(\frac{9^{1-n}-1}{1-n}\right)B_{2-2n}\left[\pi\,\mathcal{I}_0\left(\frac{3}{4}\right)\right]^{-n} \, \ ,\\[0.5em]
& \mathcal{I}_{n}\left(\frac{1}{2}\right) = \frac{2}{3}\mathcal{I}_0\left(\frac{1}{2}\right)\left(\frac{2}{3}\right)^{-n}\left(\frac{2^{1-2n}+2^{3-4n}-1}{1-n}\right)B_{2-2n}\left[\pi\,\mathcal{I}_0\left(\frac{1}{2}\right)\right]^{-n}\, \ , \\[0.5em]
& \mathcal{I}_{n}\left(\frac{1}{4}\right) = \frac{3}{5}\mathcal{I}_0\left(\frac{1}{4}\right) \left(\frac{72}{5}\right)^{-n} \left(\frac{6+3\cdot4^{n}+2\cdot9^{n}-36^{n}}{1-n}\right)B_{2-2n} \left[\pi\, \mathcal{I}_0\left(\frac{1}{4}\right)\right]^{-n}\, \ ,
\end{align}
\label{InasI0}
\end{subequations}
where $n \leq 0$ and $B_{n}$ denotes the Bernoulli numbers. Then, using the relations \eqref{omegaLeading} as well as the expressions \eqref{InasI0}, we find that the  generating function \eqref{defG} for the coefficients $\omega^{(\ell)}_{0,n}(s_{\alpha})$ can always be rewritten in the form 
\begin{align}
\label{G0step1}
G^{(0)}(s_{\alpha},x) = 4x\left[1-\exp\left(\sum_{n=0}^{\infty}c_n(s_{\alpha})\left(\frac{\mathcal{I}_0(s_{\alpha})}{x}\right)^{2n+1}\right)\right]\, \ ,
\end{align}
where the coefficients $c_n$ depend in a non trivial way on $s_{\alpha}$.  For example for $s_{\alpha}=\frac{3}{4}$ we find
\begin{align}
c_{n}\left(\frac{3}{4}\right) = -\frac{1}{2n+1}\frac{3^{2n+1}}{4^{n+1}}\left(\frac{9^{n+1}-1}{n+1}\right)B_{2n+2} \, \ .
\end{align}
Importantly, we notice that for $s_{\alpha}=\frac{3}{4},\frac{1}{2},\frac{1}{4}$ the series appearing in \eqref{G0step1} is Borel summable. Therefore, exploiting the properties of the Bernoulli numbers, we analytically obtain the generating functions
\begin{subequations}
\begin{align}
& G^{(0)}\left(\frac{3}{4},x\right) = 4x\left[1- \text{exp}\left(-2 \int_0^{\infty} \frac{dt}{t}\frac{\sinh(-\frac{2\pi t}{3x})}{1+\cosh(-\frac{2\pi t}{3x})}\,\text{e}^{-t}\right)\right]\,  ,\\[0.5em]
&  G^{(0)}\left(\frac{1}{2},x\right) = 4x\left[1-\exp\left(-\frac{1}{2}\int_0^{\infty} \frac{dt}{t}\left(2+\text{sech}\left(-\frac{\pi t}{2x}\right)\right)\tanh\left(-\frac{\pi t}{4x}\right)\text{e}^{-t}\right)\right]\,  , \\[0.5em]
&  G^{(0)}\left(\frac{1}{4},x\right) = 4x\left[1-\exp\left(\int_{0}^{\infty} \frac{dt}{t} \, \frac{\text{sech}(\frac{\pi t}{6x})\sinh(\frac{5\pi t}{6x})}{1+2\cosh(\frac{2\pi t}{3x})}\,\text{e}^{-t}\right)\right]\,  .
\end{align}
\label{Gset}
\end{subequations}
Then we evaluate the expressions \eqref{Gset} at $x=-2\pi$ and, performing the integral over the $t$ variable, we get
\begin{subequations}
\begin{align}
& G^{(0)}\left(\frac{3}{4},-2\pi\right) = -8\pi\left(1-\frac{4\pi}{9\sqrt{3}}\right) = -\frac{4\pi}{\sqrt{3/4}}\left(\mathcal{I}_0\left(\frac{3}{4}\right)+2\sqrt{\frac{3}{4}}\right)\,  , \\[0.5em]
& G^{(0)}\left(\frac{1}{2},-2\pi\right) = -8\pi\left(1-\frac{3\pi}{8\sqrt{2}}\right) = -\frac{4\pi}{\sqrt{1/2}}\left(\mathcal{I}_0\left(\frac{1}{2}\right)+2\sqrt{\frac{1}{2}}\right)\,  , \\[0.5em]
& G^{(0)}\left(\frac{1}{4},-2\pi\right) = -8\pi\left(1-\frac{5\pi}{18}\right) = -\frac{4\pi}{\sqrt{1/4}}\left(\mathcal{I}_0\left(\frac{1}{4}\right)+2\sqrt{\frac{1}{4}}\right)\,  .
\end{align}
\label{G0data}
\end{subequations}
Based on the above analytic results we recognize a clear pattern and, therefore, it is natural to conjecture that for  generic $s_{\alpha}$ the leading order generating function is given by
\begin{align}
\label{G0conjecture}
& G^{(0)}(s_{\alpha},-2\pi) = -\frac{4\pi}{\sqrt{s_{\alpha}}}\left(\mathcal{I}_0(s_{\alpha})+2\sqrt{s_{\alpha}}\right)\,  .
\end{align}
Finally, inserting this expression in \eqref{GxyGxL} we find
\begin{align}
\label{Gxy}
G^{(0)}(s_{\alpha},-2\pi,-2\pi) = \frac{\pi}{s_{\alpha}}\mathcal{I}_0(s_{\alpha})^2-4\pi\,.
\end{align}

\begin{figure}
\centering
\includegraphics[scale=0.35]{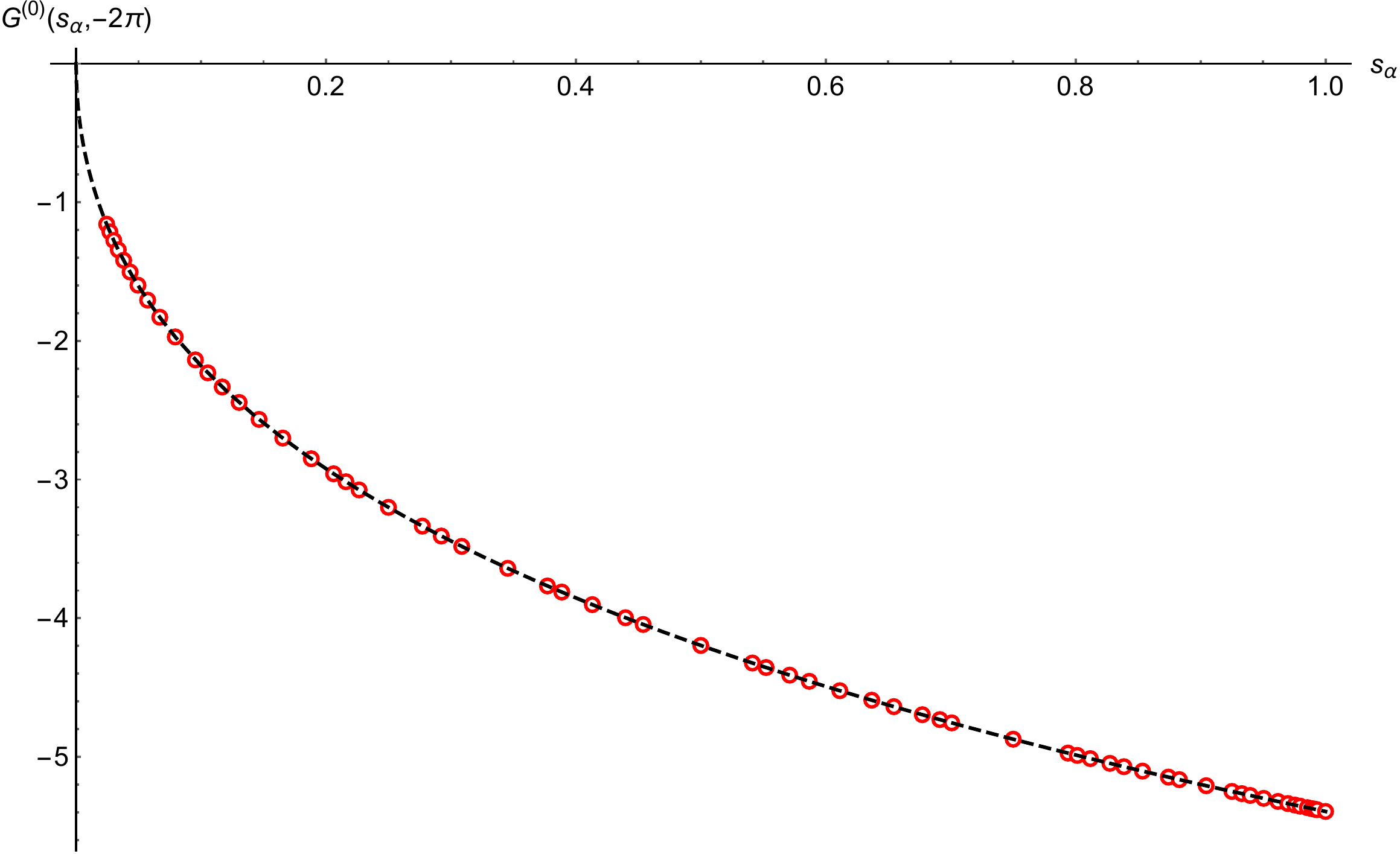}
\caption{We reported the comparison between the numerical evaluation of the function \eqref{G0conjecture} for $s_{\alpha} \in [0,1]$ (black dashed line) and its numerical estimation via a Padé resummation of the perturbative series  for fixed values of $s_{\alpha}$ (red open circles) up to $M=20$. We observe that all the points lie on the theoretical curve.}%
\label{fig:check1}
\end{figure}

\begin{figure}
\centering
\includegraphics[scale=0.35]{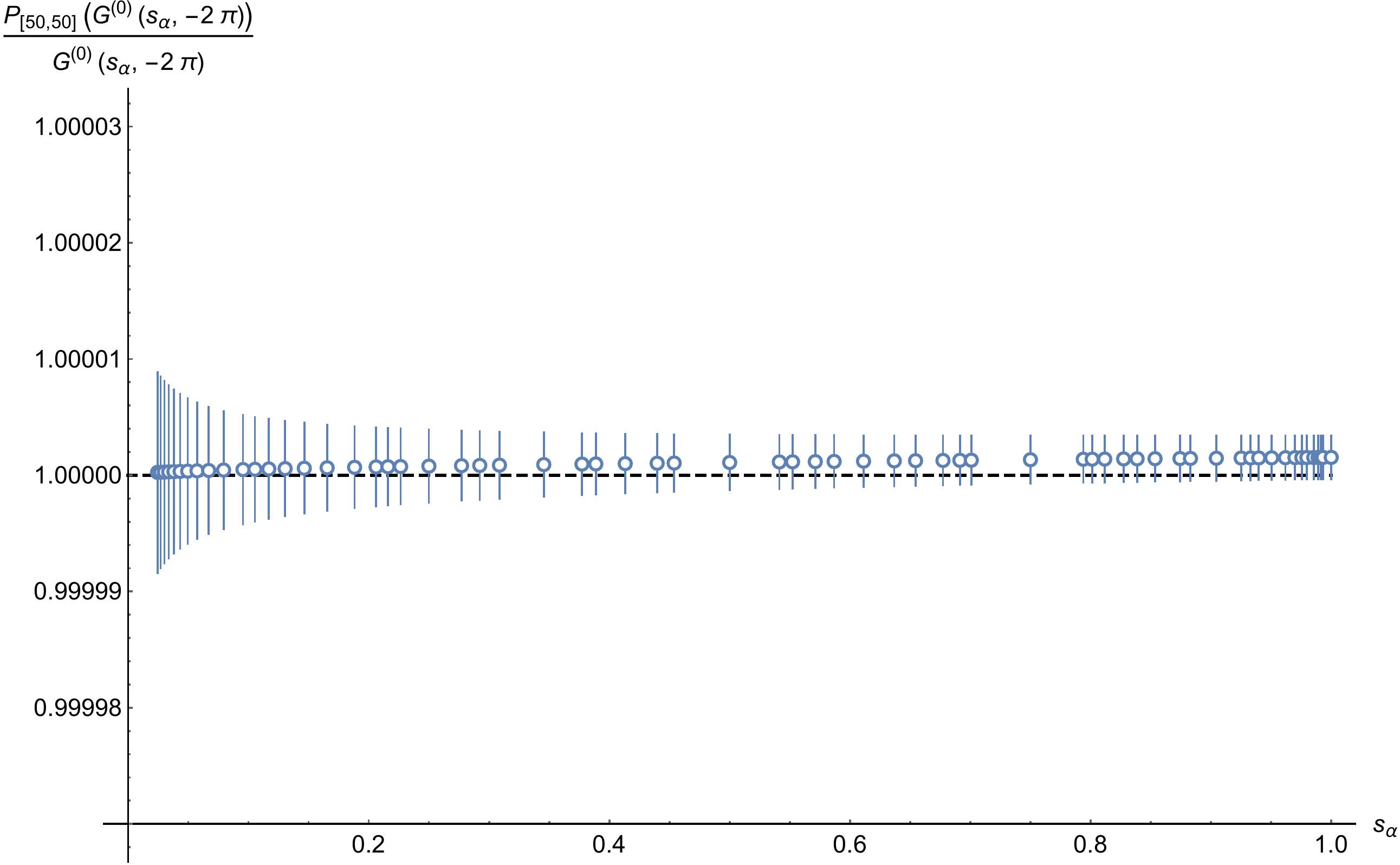}
\caption{We reported the ratios (blue open circles) among the Padé points and the function \eqref{G0conjecture} for all the different values of $s_{\alpha}$ up to $M=20$. Importantly we observe that all these ratios are compatible with the value one within numerical errors.}
\label{fig:check2}
\end{figure}

Probably a formal proof of the expression \eqref{G0conjecture} can be worked out on a case by case basis for each different value of $s_{\alpha}$. However, based on the above analysis, we have not identified a clear strategy that could lead us to proof \eqref{G0conjecture} in full generality. For this reason we choose to validate the expression \eqref{G0conjecture} numerically via a Padé resummation of the corresponding series\footnote{We refer the reader to Section \ref{subsec:checknumZ2} for an explanation of this numerical technique.}. 
As a matter of fact, exploiting the relations \eqref{wodd}-\eqref{weven} and computing numerically the integrals $\mathcal{I}_{n}(s_{\alpha})$, we evaluated the coefficients $\omega_{0,n}(s_{\alpha})$ up to $n=102$ and hence, using these data, we provide an estimation of the generating function $G^{(0)}(s_{\alpha},x)$ with a diagonal Padé  of degree $50$, namely
\begin{align}
\label{PadeG0}
P_{[50/50]}\bigl(G^{(0)}(s_{\alpha},x)\bigr)=\left[\sum_{n=0}^{102}\frac{\omega_{0,n}(s_{\alpha})}{x^n}\right]_{[50/50]}\,.
\end{align}
Then we evaluate it at $x=-2\pi$.  We collect the results of this computation for the first values of $s_{\alpha}$ in Table \ref{tab:Galpha}.
\begin{table}[ht!]
\centering
\begin{tabular}{c|c}
 $s_{\alpha}$    &  $P_{[50/50]}(G^{(0)}(s_{\alpha},-2\pi))$ \\[0.5em]
    \hline\\
     $\frac{1}{4}$    & -3.2002(8) \\[0.5em]
     $\frac{5-\sqrt{5}}{8}$ & -3.6400(4) \\[0.5em]
     $\frac{1}{2}$ & -4.1961(4)\\[0.5em]
     $\frac{3}{4}$ & -4.8724(4) \\[0.5em]
     $\frac{5+\sqrt{5}}{8}$ & -5.2079(1) \\[0.5em]
     1 & -5.3935(4) \\
    \end{tabular}
    \caption{Numerical evaluation of the generating function $G^{(0)}(s_{\alpha},-2\pi)$ for the first values of $s_{\alpha}$ via a Padé resummation of the corresponding series.}
    \label{tab:Galpha}
\end{table}
As a first check of this numerical analysis we observe that our prediction for $s_{\alpha}=1$ (last row of Table \ref{tab:Galpha}) is in remarkable good agreement with the results of \cite{Beccaria:2023kbl}, where it was conjectured that
\begin{equation}
\label{G0analytical}
G^{(0)}(1,-2\pi) =  8\pi \left[\frac{\Gamma^2(\frac{1}{2}-\frac{x}{2\pi})}{\Gamma^2(-\frac{x}{2\pi})}+\frac{x}{2\pi}\right]_{x=-2\pi} = -8\pi\left(1-\frac{\pi}{4}\right) \simeq -5.39353 \, \ .
\end{equation}
Furthermore, the numerical results found for $s_{\alpha}=\frac{3}{4},\frac{1}{2},\frac{1}{4}$ are in agreement with the theoretical expressions \eqref{G0data}. These checks permited us to test the accuracy of the Padé resummation \eqref{PadeG0} and strongly suggest that we can rely on this technique also for other values of $s_{\alpha}$. Therefore we performed a numerical evaluation of the generating function $G^{(0)}(s_{\alpha},x)$ at $x=-2\pi$ for all the values of $s_\alpha$ up to $M=20$, the results of this analysis are reported in Figure \ref{fig:check1} and Figure \ref{fig:check2}. We observe that all the numerical points are compatible with the theoretical prediction within numerical errors. We regard these numerical results as a strong confirmation of our theoretical prediction \eqref{G0conjecture}.  

\newpage

\section{Evaluation of \texorpdfstring{$\Phi^{(j)\,\text{odd}}_L(x)$}{} and \texorpdfstring{$\Phi^{(j)\,\text{even}}_L(x)$}{}}
\label{app:phi}

Here we evaluate  the functions $\Phi_L^{(j)\,\text{odd}}$ and $\Phi_L^{(j)\,\text{even}}$ defined in Section \ref{sec:2pt} for the first values of $L$, in order to understand how to rewrite the 2-point correlator, and then all the $n$-point functions, in terms of the $w_{n,m}^{(\ell)}(s_{\alpha})$ coefficients. 

For $\Phi_L^{(1)\,\text{odd}}(x)$ we find
\begin{subequations}
\begin{align}
& \Phi^{(1)\,\text{odd}}_0(x) = -\frac{1}{2}J_{2}(\sqrt{x})\,,\\
& \Phi^{(1)\,\text{odd}}_1(x) = \left[1+(x\partial_x)\right]J_{2}(\sqrt{x})\,,\\
& \Phi^{(1)\,\text{odd}}_2(x) = \left[\frac{1}{2}-\frac{3}{2}(x\partial_x)-2(x\partial_x)^2\right]J_{2}(\sqrt{x})\,,\\
& \Phi^{(1)\,\text{odd}}_3(x) = \left[\frac{3}{8}-\frac{29}{8}(x\partial_x) + 4(x\partial_x)^3\right]J_{2}(\sqrt{x})\,, \\
& \Phi^{(1)\,\text{odd}}_4(x) = \left[\frac{3}{8}-\frac{77}{8}(x\partial_x) +8(x\partial_x)^2+10(x\partial_x)^3-8(x\partial_x)^4\right]J_{2}(\sqrt{x}) \,.
\end{align}
\end{subequations}
For $\Phi^{(2)\,\text{odd}}_L(x)$ we find
\begin{subequations}
\begin{align}
& \Phi^{(2)\,\text{odd}}_0(x) = -\frac{1}{2}J_2(\sqrt{x})\,, \\
& \Phi^{(2)\,\text{odd}}_1(x) =\left[\frac{1}{4}+ (x\partial_x)\right]J_2(\sqrt{x})\,, \\
& \Phi^{(2)\,\text{odd}}_2(x) = \left[\frac{17}{16}-2(x\partial_x)^2\right]J_2(\sqrt{x})\,, \\
& \Phi_3^{(2)\,\text{odd}}(x) = \left[\frac{33}{16}-4(x\partial_x)-3(x\partial_x)^2+4(x\partial_x)^3\right]J_2(\sqrt{x})\,,  \\
& \Phi_4^{(2)\,\text{odd}}(x) = \left[\frac{825}{256}-16(x\partial_x)+\frac{17}{4}(x\partial_x)^2+16(x\partial_x)^3-8(x\partial_x)^4\right]J_2(\sqrt{x}) \,.
\end{align}
\end{subequations}
For $\Phi_{L}^{(1)\,\text{even}}$ we find
\begin{subequations}
\begin{align}
& \Phi_0^{(1)\,\text{even}}(x) = -\frac{1}{2}J_{1}(\sqrt{x})\,, \\
& \Phi_1^{(1)\,\text{even}}(x) = \left[\frac{1}{4}+(x\partial_x)\right]\,J_1(\sqrt{x})\,, \\
& \Phi_2^{(1)\,\text{even}}(x) = \left[\frac{5}{16}-2(x\partial_x)^2\right]J_1(\sqrt{x})\,,\\
&\Phi_3^{(1)\,\text{even}}(x) = \left[\frac{9}{16}-(x\partial_x)-3(x\partial_x)^2+4(x\partial_x)^3\right]J_1(\sqrt{x})\,, \\
& \Phi_4^{(1)\,\text{even}}(x) = \left[\frac{369}{256}-4(x\partial_x)-\frac{19}{4}(x\partial_x)^2+16(x\partial_x)^3-8(x\partial_x)^4\right]J_1(\sqrt{x})\,.
\end{align}
\end{subequations}
For  $\Phi_{L}^{(2)\,\text{even}}$ we find
\begin{subequations}
\begin{align}
& \Phi_0^{(2)\,\text{even}}(x) = -\frac{1}{2}J_{1}(\sqrt{x})\,, \\
& \Phi_1^{(2)\,\text{even}}(x) = \left[-\frac{1}{2}+(x\partial_x)\right]\,J_1(\sqrt{x})\,, \\
& \Phi_2^{(2)\,\text{even}}(x) = \left[-\frac{1}{4}+\frac{3}{2}(x\partial_x)-2(x\partial_x)^2\right]J_1(\sqrt{x})\,,\\
&\Phi_3^{(2)\,\text{even}}(x) = \left[\frac{9}{16}+\frac{7}{8}(x\partial_x)-6(x\partial_x)^2+4(x\partial_x)^3\right]J_1(\sqrt{x})\,, \\
& \Phi_4^{(2)\,\text{even}}(x) = \left[\frac{45}{16}-\frac{29}{8}(x\partial_x)-13(x\partial_x)^2+22(x\partial_x)^3-8(x\partial_x)^4\right]J_1(\sqrt{x})\,. 
\end{align}
\end{subequations}

\printbibliography

\end{document}